\documentclass[preprint,12pt]{elsarticle}
\usepackage{siunitx,microtype,amsmath,amssymb,nicefrac,float,color,hyperref}
\usepackage{ltablex,booktabs}
\usepackage[tableposition=below]{caption}
\usepackage[square,numbers]{}

\definecolor{purp}{rgb}{0.4,0.2,0.8}
\definecolor{green}{rgb}{0.255,0.57,0.295}

\newcommand{\R}{\mathbb{R}}

\newcommand{\vv}{\vec{v}}
\newcommand{\vy}{\vec{y}}

\newcommand{\vq}{\vec{q}}
\newcommand{\vqd}{\dot{\vec{q}}}
\newcommand{\vqdd}{\ddot{\vec{q}}}
\newcommand{\dcon}{d_\text{con}}
\newcommand{\Atol}{\textit{Atol}}
\newcommand{\Rtol}{\textit{Rtol}}
\renewcommand{\vec}[1]{\mathbf{#1}}
\newcommand{\Trans}{\mathsf{T}}

\begin{document}
\title{Unravelling the Mechanics of Knitted Fabrics Through Hierarchical Geometric Representation}

\author[inst1,inst2]{Xiaoxiao Ding}
\affiliation[inst1]{organization={Harvard John A. Paulson School of Engineering and Applied Sciences, Harvard University, Cambridge, MA 02138, USA}} \ead{xiaoxiaoding@g.harvard.edu}
\affiliation[inst2]{organization={Department of Mathematics, University of Wisconsin--Madison, Madison, WI 53706, USA}}

\author[inst1,inst3]{Vanessa Sanchez}
\affiliation[inst3]{organization={Department of Chemical Engineering, Stanford University, 443 Via Ortega, Stanford, CA 94305, USA}}

\author[inst1]{Katia Bertoldi}

\author[inst2,inst4]{Chris H. Rycroft}
\affiliation[inst4]{organization={Computational Research Division, Lawrence Berkeley Laboratory, 1 Cyclotron Road, Berkeley, CA 94720, USA}}
\ead{chr@math.wisc.edu}

%%%% Subject entries to be placed here %%%%
%\subject{applied mathematics, computational physics, continuum mechanics}

%%%% Keyword entries to be placed here %%%%
%\keywords{nonlinear elasticity, hierarchical geometry, topology, knitted fabrics, dynamical simulation, mechanical programmability}

%%%% Abstract text to be placed here %%%%%
\begin{abstract}
Knitting interloops one-dimensional yarns into three-dimensional fabrics that exhibit behaviour beyond their constitutive materials. How extensibility and anisotropy emerge from the hierarchical organisation of yarns into knitted fabrics has long been unresolved. We seek to unravel the mechanical roles of tensile mechanics, assembly and dynamics arising from the yarn level on fabric nonlinearity by developing a yarn-based dynamical model. This physically validated model captures the mechanical response of knitted fabrics, analogous to flexible metamaterials and biological fiber networks due to geometric nonlinearity within such hierarchical systems. Fabric anisotropy originates from observed yarn--yarn rearrangements during alignment dynamics and is topology-dependent. This yarn-based model also provides a design space of knitted fabrics to embed functionalities by varying geometric configuration and material property in instructed procedures compatible to machine manufacturing. Our hierarchical approach to build up a knitted fabrics computationally modernizes an ancient craft and represents a first step towards mechanical programmability of knitted fabrics in wide engineering applications.
\end{abstract}
%%%%%%%%%%%%%%%%%%%%%%%%%%%%%%%%%%%%%%%%%%%%%%%%

%%%% End of first page %%%%%%%%%%%%%%%%%%%%%
\maketitle

%%%% Section 1 %%%%%%%%%%%%%%%%%%%%%%%%
\section{Introduction}
\label{sec:intro}
%%%% Section 1 %%%%%%%%%%%%
\label{sec:1}

Knitted fabrics are hierarchical structures that build up from yarns at the microscale, to stitch pattern structures at the mesoscale, and finally to three-dimensional fabrics at the macroscale. With yarns being the primary building blocks to dominate the physics and design of fabrics, the separation of these two scales (i.e., yarn-level and fabric-level) at the structural level not only causes a range of interesting physical phenomena \cite{Poincloux2018, Poincloux2018a} to arise, but also provides a huge design space for functionalities in knitted fabrics beyond what their constitutive materials can achieve \cite{Rout2022}, echoing a wide range of research interests to design functional materials. Knitted fabrics, analogous to some architected \cite{Compton2014, zheng14, Jiang2016, Moestopo2023} and bio-inspired materials \cite{Gladman2016, Ma2017, Nepal2023, Mistry2023}, represent nonbiological examples of a nonlinear elastic response characterised by a ``J-shape'' curve, as the fabric transitions from bending energy dominant region to stretching energy dominant region under uniaxial tension. The presence of mesoscale stitch patterns enables the fabric to take on substantial tensile stress elastically. This behaviour is attributed to the low strains on individual yarn segments and the dynamics of yarn alignment with external load, which offer geometric degrees of freedom. The distinctively compliant behaviour of knitted fabrics makes them stand out as excellent scaffolds for wearable devices \cite{Polygerinos2015, Cappello2018, lee2018knit, Granberry2019, fan2020machine, Wicaksono2020} and soft robotic actuators \cite{Connolly2019,Sanchez2021}, where large deformation and flexible morphing without material damage is desired. The anisotropy at the fabric mesoscale has been exploited to fine tune actuation of such devices \cite{Ahlquist2017,Luo2022,Sanchez2023}, where carefully selected structures can be spatially varied across the fabric, such that the fabric can shape morph to comply with complex geometries. Multifunctional knitted fabrics can be created through embedding functional yarns into conventional knit structures, to further enlarge the design space of knitted fabrics to morphing structures \cite{Abel2013, Han2017} and to serve as light and touch sensors \cite{Albaugh2019}, pressure sensors \cite{Luo2021}, electronic interfaces \cite{Wicaksono2017} and electronic skins \cite{Pei2019} in an exciting new domain of smart materials to mimic and embed intelligence. 

Currently, intuition-led strategies remain the primary approach to design devices made of knitted fabrics. This paradigm poses limitations on exploring the design space due to high machinery costs, training costs and material waste. A generalisation of these application-driven designs for yarn geometries, fabric structures and material variations has not yet been established. Early theoretical work on knitted structures started from defining the characteristic unit cells to represent stitch patterns and predominantly assumed homogeneity due to periodicity of unit cells. Starting from a three-dimensional parameterisation of the jersey knit pattern \cite{Peirce1947} to curvature augmented model \cite{Leaf1955}, followed by energy minimisation model \cite{Munden1959}, most geometric models of knitted fabrics are constructed through superposition of cosine and sine curves due to the smoothness and periodicity of these shape functions. With the development of spline basis functions, we can discretise yarns with sufficient accuracy and such yarn-based models \cite{Bergou2008, Kaldor2008, Cirio2014, Liu2017, Leaf2018, Wu2020} have key advantages compared to coarse-grained models \cite{Terzopoulos1987, Baraff1998, Breen1994, Yeoman2010} and homogenised models \cite{NarainArminSamiiJamesO2012, Yuksel2012, Dinh2018, Weeger2018, Sperl2021}, due to their capability to (i) capture mechanical behavior originating from first principles via yarn dynamics, (ii) provide quantitative measurements of geometric nonlinearity arising across scales, and (iii) vary the spatial distribution of stitch patterns and material properties of yarns to form targeted 2D and 3D configurations, all while not constraining the extensibility of individual yarn segments affinely.

To begin, we adopt a yarn-based model with cubic spline basis functions \cite{Kaldor2008} that was originally applied in computer graphics to animate cloth in a qualitatively realistic manner. We extend this model to provide physical insights into macroscopic inhomogeneity, anisotropy and cross-scale mechanisms. We investigate the mechanical responses of representative weft-knitted samples at compatible scale to machine knitted experimental samples that are systematically characterised. Our numerical approach is implemented through fully dynamic formulation of the governing equation of motion at yarn level, integrated explicitly with a high-order adaptive scheme. A key aspect of our numerical procedure is the introduction of relaxation stage similar to experimental procedure \cite{Martinez2021} before the application of external tensile forces, to account for residual stress that is inherent in the knit fabrication process. This inherent residual stress has been experimentally investigated by the textile community as one of the dimensional properties of knitted fabrics \cite{Allan1983, Amreeva2007, Wei2011} and poses a typical challenge in generating accurate reference state of knitted fabric, if obtained purely from geometric description of its structure. After initialisation, we apply uniaxial tensile loads quasi-statically to stretch the fabric samples in simulation, up to strain thresholds compatible with experimental set-up post initial cycles. Our designed experimental procedure is compatible with bias-extension tests typically conducted to characterise textiles \cite{Cao2008}. Though experimental validation is carried out on limited sets of weft-knitted samples, our selection of stitch pattern each represents a different topological group \cite{Markande2020} that can be further explored in a more systematic approach. 

Leveraging parameterisation at the yarn level, we can quantitatively investigate cross-scale mechanisms contributing to nonlinear elasticity and anisotropy of knitted fabrics. Since we treat the dynamics at a continuous yarn as the governing mechanics, allowing for the prediction of local, spatial evolution of bending and stretching energy, we can explain mechanical responses of representative stitch patterns by statistical measurements of yarn dynamics. As we can directly compute measurements of energy, deformation and alignment with regard to each yarn segment, we can predict micromechanical hot spots. 

Another main aspect of the present work is to propose design of functional and composite textiles in an instructed manner compatible to manufacturing procedure, allowing our model to be adapted for systematic digital generation of knitted configurations for targeted mechanical responses. Guided by gained physical insights, we can purposely design 2D and 3D configurations that possess localised mechanical responses through spatial distribution of stitch structures and material properties. The direct applications of mechanically programmed knitted fabrics in responsive structures, wearables and soft robotics are emphasised in this work. Though demonstrated design examples focus on one length scale, our model can also be deployed to various length scales and other fibrous networks.

More broadly, the textiles that we consider can be viewed part of the broader class of mechanical metamaterials~\cite{barchiesi19} and programmable materials~\cite{florijn14}, which have attracted much attention recently for their ability to access new functionality such as high strength to weight ratios~\cite{zheng14} or negative Poisson's ratio~\cite{grima13}. Many of the same research questions, such as in fabrication~\cite{duoss19,yuan19}, as well as exploration~\cite{medina20,chan20} and optimization~\cite{mao20,deng22,medina23} of the design space, can potentially be extended to the textiles that we consider.

This paper is organised as follows: we first summarise theoretical background of the geometric model and the mechanical model; in addition to providing validated numerical results and motivating examples for applications, we demonstrate that structural properties, such as varying topological description of fabric patterns and varying spatial distribution of fabric patterns can effectively adjust the mechanical behaviours of knitted fabrics, not necessarily modifying the material properties in results; moreover, we detail the development and implementation of computational model, the material selection and characterisation in following sections. 
\vskip6pt

%%%% Section 2 %%%%%%%%%%%%%%%%%%%%%%%%
\section{Theoretical background}
\label{sec:theory}
%%%% Section 2 %%%%%%%%%%%%
\subsection{Discretisation of the yarn, and initial conditions for the fabric}
\label{sec:2a}
Our work builds on previous yarn-level simulations in the computer graphics literature by Kaldor et al.~\cite{Kaldor2008}. We implemented a custom code in C++, using OpenMP for multithreading, to perform the simulations. In this section we provide a mathematical overview of the methods, and then provide additional technical and computational details in ~\ref{sec:5}.

Our simulation can handle an arbitrary number of individual yarns of diameter
$d$. The centerline of each yarn is represented by a cubic B-spline with $N$
segments. We denote $s\in \Omega = [0,N]$ to be a dimensionless coordinate
along the yarn, where the $N+1$ control points of the spline are located at
$s=0,1,2, \ldots, N$. The cubic B-spline basis function is given by
\begin{equation}
  B(s) = \begin{cases}
    \tfrac23 + s^2(\tfrac12 |s|-1) & \qquad \text{if $|s|<1$,} \\
    \tfrac{1}{6}(2-|s|)^3 & \qquad \text{if $1\le |s| < 2$,} \\
    0 & \qquad \text{otherwise.}
  \end{cases}
\end{equation}
A family of basis functions is then given by $b_k(s) = B(s-k)$ for $k\in \mathbb{Z}$. The yarn
is then defined as
\begin{equation}
  \vy(s,t) = \sum_{k=-1}^{N+1} b_k(s) \vq_k(t),
  \label{eq:ys}
\end{equation}
where $\vq_k(t)$ are time-dependent three-dimensional functions. In
Eq.~\eqref{eq:ys}, the sum must run from $-1$ to $N+1$ in order to describe all
piecewise cubics in $C^2[0,N]$~\cite{suli_textbook}, making for $m=N+3$ terms
in total. Hence, the yarn is described by $3m=3(N+3)$ degrees of freedom stored
in a vector $\vq=(\vq_{-1},\vq_0,\ldots,\vq_{N+1})$. The velocity of the yarn
is given by
\begin{equation}
  \vv(s,t) = \dot{\vec{y}}(s,t) = \sum_{k=-1}^{N+1} b_k(s) \vqd_k(t),
  \label{eq:vs}
\end{equation}
where a dot represents a derivative with respect to $t$. The velocity of the
yarn is analogously described by a $3m$-component vector $\vqd$. The descriptions
in Eqs.~\eqref{eq:ys} \& \eqref{eq:vs} effectively decouple the spatial
and temporal dependence of the yarn motion. In its rest state, the yarn has
equal arc length $l$ between each pair of control points.

We initialise the spline at $t=0$ by specifying an initial parametric curve for its shape. Knitted fabrics are generated from interlocking loop units that are formed stitch by stitch in the weft
(horizontal) and warp (vertical) directions. Depending on the direction along
which a continuous yarn is fed in, knitted fabrics fall into two categories: weft knits and warp knits. We focus on weft-knitted fabrics in this work, because of the current interest in leveraging commercially available weft knitting machines (V-bed knitting machines) to create complex 3D devices.

A typical loop geometry that we employ is~\cite{Vassiliadis2007}
\begin{equation}
  \vy_\text{p}(w) = \left(
    \begin{array}{c}
      \lambda_x \Big( w+\sin(\pi w) \Big) \\
      \lambda_y \cos(\frac{\pi w}{2}) \\
      \lambda_z \cos(\pi w)
    \end{array}
  \right)
  \label{eq:basic_param}
\end{equation}
where $\lambda_x, \lambda_y, \lambda_z$ are scaling parameters in each
dimension that may vary independently to match target aspect ratio of generated
samples. Using the coordinate range $w\in
[w_\text{start},w_\text{end}]=[-\tfrac{\pi}{2},\tfrac{\pi}{2}]$ in
Eq.~\eqref{eq:basic_param} yields a single loop as shown in
Fig.~\ref{fig1:knit_assembly}(A). In general, $\|d\vy_\text{p}/dw\|$ will not be
constant, so that the arc length along each parametric curve will not increase
at a constant rate in $w$. Therefore, to initialize the B-spline formulation,
our simulation computes the arc length along the curve as a function of $w$,
\begin{equation}
  A(w) = \int_{w_\text{start}}^w \left| \frac{d\vy_\text{p}}{dw} \right| dw,
\end{equation}
which is evaluated using composite Gaussian quadrature. The rest arc length is
computed as $l= A(w_\text{end})/N$. Using Ridders' root-finding method, a
sequence of values $w_0=w_\text{start}, w_1, w_2,\ldots, w_N=w_\text{end}$ are
found such that $A(w_k) = kl$. These set the values of the control points in
the B-spline basis formulation, so that $\vy(k,0) =\vy_\text{p}(w_k)$, giving
$N+1$ vector equations in total. In addition, the direction of the spline at
$s=0,N$ is chosen to match the direction of the parametric curve, giving an
additional two vector equations. This gives a total of $N+3$ linear vector
equations that can be solved as a linear system to determine the $\vq_k(0)$.

\subsection{Dynamics of a yarn}
\label{sec:2b}
We extend the Lagrangian formulation to describe the dynamics of a single
yarn with $m$ control points as
\begin{equation}
  \frac{d}{dt} \left( \nabla_{\vqd_k} T \right) + \nabla_{\vq_k} V + \nabla_{\vqd_k} D =0,
  \label{eq:lagr}
\end{equation}
where $T$ is the kinetic energy, $V$ is the potential energy and $D$ is the damping term. The kinetic energy of the yarn is
\begin{equation}
  T(\vqd) = \frac{\rho l}{2} \int_\Omega \vv^\Trans \vv \, ds,
\end{equation}
where $\rho$ is the mass density. By referencing Eq.~\eqref{eq:lagr}, we must
evaluate
\begin{equation}
  \nabla_{\vqd_k} T = \rho l \int_\Omega (\nabla_{\vqd_k} \vv^\Trans) \vv \, ds.
  \label{eq:dkint}
\end{equation}
We define the unit mass matrix $M\in \R^{m\times m}$ with components
\begin{equation}
  M_{jk} = \int_\Omega b_k(s) b_j(s) \, ds,
  \label{eq:ke_integ}
\end{equation}
which corresponds to integrating a product of two B-spline basis functions
$b_k$ and $b_j$. Since each basis function is non-zero over four intervals,
$M_{jk}=0$ if $|k-j|>3$, and therefore $M$ is a banded matrix with three
superdiagonals and three subdiagonals. The matrix $M$ remains constant throughout the
simulation and can be precomputed. Therefore Eq.~\eqref{eq:dkint} becomes
\begin{equation}
  \frac{d}{dt} \left( \nabla_{\vqd_k} T \right) = \rho l \sum_{k=-1}^{N+1} M_{jk} \vqdd_{j}.
    \label{eq:Mqdd}
\end{equation}
The potential energy of a yarn includes several terms as
\begin{equation}
    V = V^s(\vq)+V^b(\vq)+V^g(\vq),
    \label{eq:pot_en}
\end{equation}
representing energy due to stretching, bending, and gravity. With the assumption of linear elasticity, the stretching energy is given by
\begin{equation}
	V^{s}(\vq) = \frac{E^s A l}{2} \int_\Omega \Big( \frac{\|\Vec{y'}\|}{l}-1 \Big) ^2 ds,
\end{equation}
where $E^s$ is the tensile stiffness and $A=\pi d^2/4$ is the yarn cross-sectional area. Here, the prime superscript represents a partial derivative with respect to $s$. The elastic energy of the yarn due to bending is formulated as
\begin{equation} \label{eq:bend_en}
	V^{b}(\vq) = \frac{E^b I l}{2} \int_\Omega \kappa^2 ds,
\end{equation}
where $E^b$ is the bending stiffness, $I$ represents moment of inertia of the yarn cross-section, and the local curvature $\kappa$ is defined as
\begin{equation} \label{eq:curvature}
	\kappa = \frac{\|\Vec{y}' \times \Vec{y}''\|}{\|\Vec{y}'\|^3}.
\end{equation}  
The gravitational potential energy is
\begin{equation}
  V^g(\vq) = \rho l \int_\Omega \vy^\Trans \vec{g} ds,
\end{equation}
where $\vec{g}$ is the gravitational acceleration. By referencing
Eq.~\eqref{eq:lagr}, we need to evaluate $\nabla_{\vq_k} V^s(\vq)$,
$\nabla_{\vq_k} V^b(\vq)$ and $\nabla_{\vq_k} V^g(\vq)$. Similar to evaluating
Eq.~\eqref{eq:Mqdd}, part of these complicated integrals can be precomputed and
the rest can be accurately determined using quadrature.

The damping term in Eq.~\eqref{eq:lagr} has several components. One component
is given by
\begin{equation} \label{eq:damp_en_g}
	D^\text{iso}(\vqd) = k_{g} \int_\Omega \vv^\Trans \vv \, ds,
\end{equation}
which creates a global drag force on the yarns. In the experiments, the knitted
samples are primarily in a regime where the forces are in quasi-static
equilibrium, since the yarns have sufficient internal damping to remove any
transient inertial effects. The drag force in Eq.~\eqref{eq:damp_en_g}
serves as a simple proxy for the internal damping and accomplishes the same
goal, ensuring that the inertial effects are removed.

\subsection{Yarn--yarn interactions}
\label{sec:2c}
The contact forces between two yarns (or between two different sections of the same yarn) are critically important for simulating the knitted fabric. Without loss of generality, let $s$ and $\tilde{s}$ be coordinates ranging from 0 to 1 over two spline segments $i$ and $j$ with a contact. The energy contribution is given by
\begin{equation}
  V^\text{con}_{i,j} = l^2 \int_0^1 \int_0^1 f\left( \frac{\| \vy_i(\tilde{s}) - \vy_j(s) \|}{d} \right) \,ds \, d\tilde{s},
  \label{eq:contact_integ}
\end{equation}
where $\vy_i$ and $\vy_j$ are the spline positions on the two segments, and
\begin{equation}
  f(\delta) =
  \begin{cases}
    k (\delta-1)^2 & \qquad \text{if $0 \le \delta <1$,} \\
    0 & \qquad \text{if $\delta \ge 1$,}
  \end{cases}
\end{equation}
where $k$ is a spring constant to represent contact repulsive stiffness. In addition a damping term can be incorporated,
with the form
\begin{equation} \label{eq:damp_en_c}
  D_{i,j}^\text{fri} = l^2 \int_0^1 \int_0^1 \Big( k_{dt}\| \Delta\vv_{ij}\|^2 - (k_{dt}-k_{dn})(\hat{\Vec{n}}^\Trans_{ij} \Delta\vv_{ij})^2 \Big) \,ds\, d\tilde{s},
\end{equation}
which approximates the effect of frictional sliding as inter-yarn slip \cite{Syerk2012}. Here $\Delta\vv_{ij}$ is
the relative velocity and $\hat{\Vec{n}}_{ij}$ is a normal vector in the
collision direction. The constants $k_{dt}$ and where $k_{dn}$ set the size of the effect in the tangential and normal directions, respectively.

Unlike the integrals considered in the previous section, it is difficult to
evaluate the integrals in Eqs.~\eqref{eq:contact_integ} \& \eqref{eq:damp_en_c}
efficiently and accurately. Since the integrands are non-smooth, and are
only non-zero in localised patches in the $(s,\tilde{s})$ space, Gaussian
quadrature will often give imprecise results. Because of this, we replace each
integral with sum over $n$ discrete values $\{s_1,s_2,\ldots,s_n\}$ and
$\{\tilde{s}_1,\tilde{s}_2,\ldots,\tilde{s}_n\}$ so that
\begin{equation}
  V^\text{con}_{i,j} = l^2 \sum_{\alpha=1}^n \sum_{\beta=1}^n f\left( \frac{\| \vy_i(s_\alpha) - \vy_j(s'_\beta) \|}{\dcon} \right),
  \label{eq:contact_sum}
\end{equation}
where $s_\alpha=(2\alpha-1)/n$ and $\tilde{s}_\beta= (2\beta -1)/n$. This is
equivalent to modeling contact between a discrete set of spheres, evenly
distributed along each spline segment. Similarly, Eq.~\eqref{eq:damp_en_c} is
replaced with
\begin{equation}
  D^\text{fri}_{i,j} = l^2 \sum_{\alpha=1}^n \sum_{\beta=1}^n \Big( k_{dt}\| \Delta\vv_{\alpha\beta}\|^2 - (k_{dt}-k_{dn})(\hat{\Vec{n}}^\Trans_{\alpha\beta} \Delta\vv_{\alpha\beta})^2 \Big),
\end{equation}
where $\Delta \vv_{\alpha\beta}= \vv_i(s_\alpha) - \vv_j(\tilde{s}_\beta)$ and
$\hat{\Vec{n}}_{\alpha\beta}$ is a normal vector pointing in the direction of
$\vy_i(s_\alpha)-\vy_j(s_\beta)$. The diameter $\dcon$ of the contact spheres
is chosen to be slightly larger than the yarn diameter $d$, so that the
envelope made by the spheres more precisely matches the profile of the
yarn---see ~\ref{sec:csphere} for more information.

To detect the adjacent spheres efficiently, the spheres are binned into an
equally-spaced rectangular grid that covers all of the yarns in the simulation.
For a given sphere, finding adjacent spheres is performed by iterating over all
spheres in nearby grid boxes, resulting in a constant, $O(1)$ computation time
per sphere. Even with this optimisation, we typically find that detecting and
computing the contact forces is the most computationally expensive step
of our simulations.

\subsection{Loop topology and fabric pattern}
\label{sec:2d}
\begin{figure}[H]
    \centering
    \includegraphics[width=0.5\textwidth]{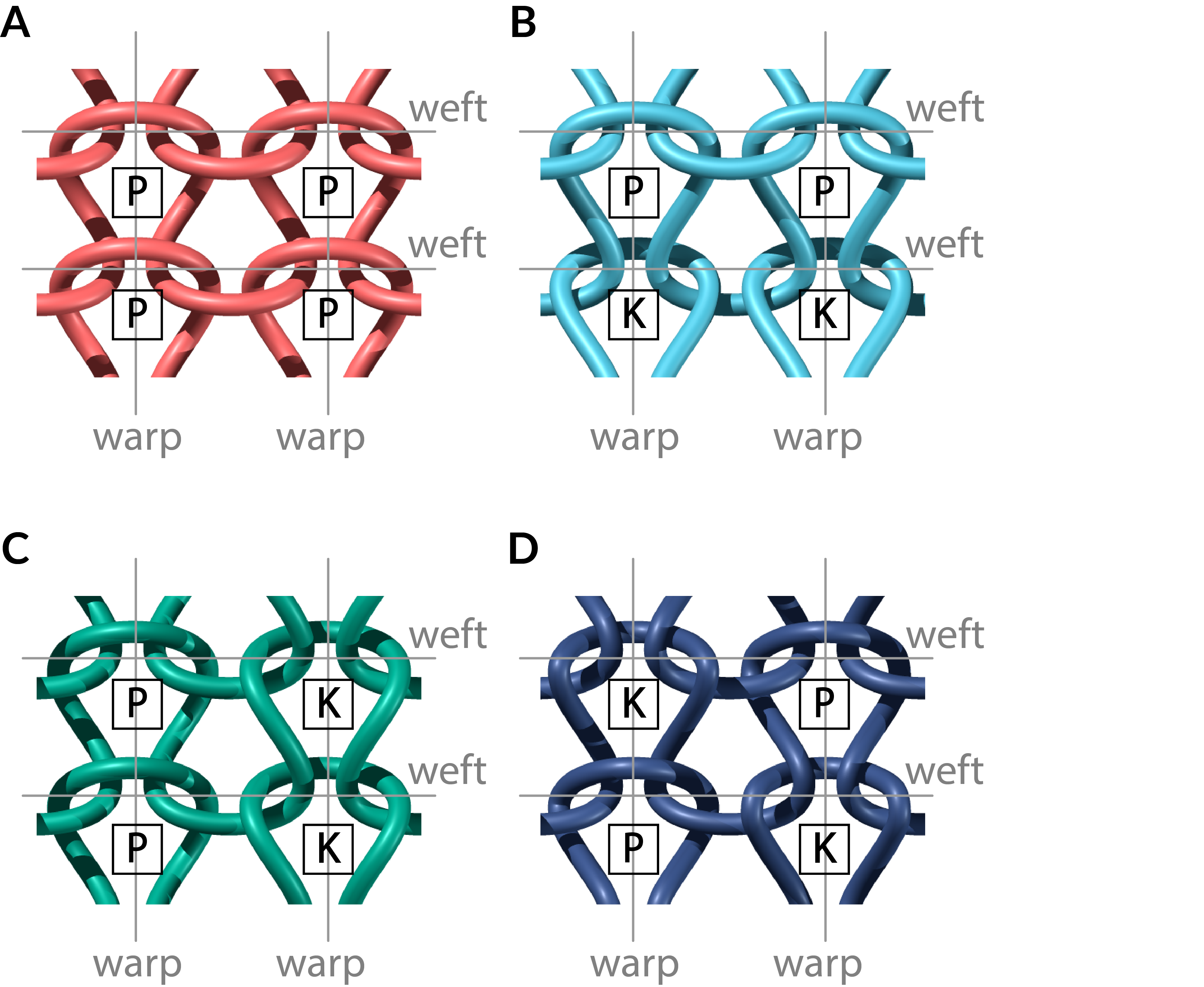} \vspace{-1em}
    \caption{Illustration of basic weft-knitted fabric patterns composed of two by two contacts with varying topology (each contact can either be purl (P) or knit (K)): (A) jersey, (B) garter 1 by 1, (C) rib 1 by 1 and (D) seed 1 by 1.\label{fig2:knit_patterns}\vspace{-1em}}
\end{figure}

A typical V-bed knitting machine consists of front and back beds with arrays of needles. A carriage traverses these beds, actuating the knitting needles with cams. Concurrently, yarn carriers (moved by the carriage in our case) feed yarns to be caught by needles to form stitches. Stitches formed on the front bed resemble ``knit'' stitches, while those on the back bed are akin to ``purl'' stitches in hand knitting. Based on these two basic manufacturing instructions, ``knit'' and ``purl,'' we can define a set of four representative weft-knit structures as shown in Fig.~\ref{fig2:knit_patterns} (A) jersey (all knits or purls), (B) garter 1 by 1 (knits and purls alternating every row only), (C) rib 1 by 1 (knits and purls alternating every column only) and (D) seed 1 by 1 (knits and purls alternating every row and column). 

\begin{figure}
    \centering
    \includegraphics[width=0.8\textwidth]{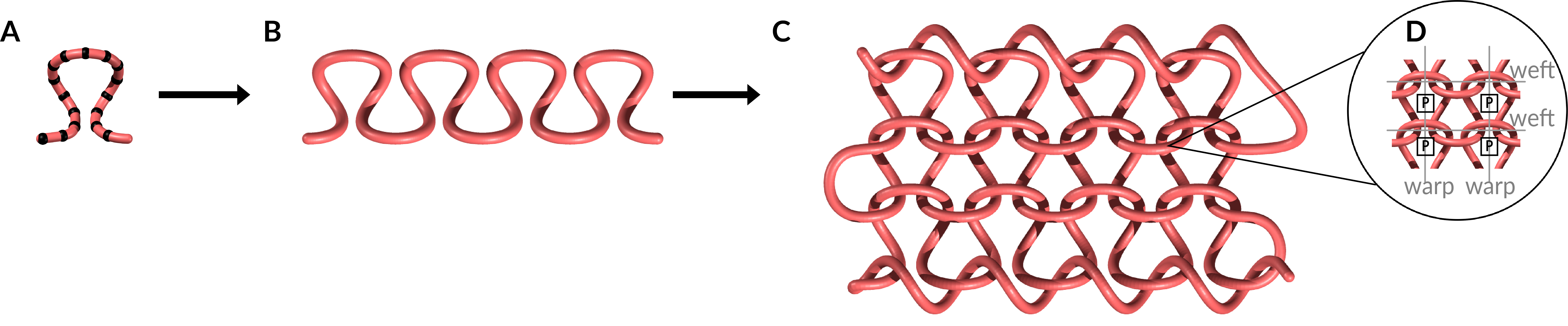}\vspace{-0.5em}
    \caption{The assembly of a representative weft-knitted fabric (jersey). (A) At the microscale: a loop discretised with evenly spaced control points along fixed cubic basis functions.  (B)  At the mesoscale: each row of yarn formed along fabric weft and translated along fabric warp based on the topological description of representative structure (We present jersey as an example here and the same assembly process is generalised for other patterns, such as garter, rib and seed). (C) At the macroscale: a fabric with ends of each row connected and additional spiral yarns attached to the bottom (``casting on'' in textile terminology) and top boundary (``binding off'' in textile terminology) to prevent fabric from unravelling. }
    \label{fig1:knit_assembly}
\end{figure}

To be consistent with the manufacturing process~\cite{Spencer2001}, we propose a simple pipeline to assemble full-scale weft knitted fabrics. Fig.~\ref{fig1:knit_assembly} illustrates the process of creating a geometric model for jersey, the simplest weft knitted fabric, since the topology of all contacts within the fabric is consistent. After generating a loop unit along the standardised parametric function, we assemble a row of loops by assigning the end positions of the row along the $x$ axis (fabric weft direction), similar to how rows of stitches are formed along horizontal needle beds on the V-bed knitting machine. Secondly, we translate each row along the $y$ axis (fabric warp direction) with assigned distance from the central axis of the pattern. Note that modifications to the geometric model are required to adjust for spacing between alternating rows and/or columns in order to create more complex configurations beyond the jersey. We proposed using an additional sinusoidal function to parameterise the $z$ direction (fabric thickness direction) in Eq.~\eqref{eq:basic_param}, in order to alternate wavelength and apply a phase shift to accommodate for varying topology~\cite{Markande2020}. In addition, we specify smooth spiral curves adopting a generalised helicoid surface~\cite{Piuze2011, Wadekar2020} equivalent to extra yarns used to cast on and bind off the fabric at the top and bottom boundaries in manufacturing, in order to prevent fabric from unravelling upon free boundary conditions. These spiral curves can be described by 
\begin{equation}
  \vy_\text{s}(w) = \left(
    \begin{array}{c}
      \lambda_{x,\text{sp}} (0.25 w) \\
      \lambda_{y,\text{sp}} \sin \Big( \gamma \pi (w+d) \Big) \\
      \lambda_{z,\text{sp}} \Big( 1 - \cos(0.5\pi (w+d))
    \end{array}
  \right),
  \label{eq:basic_param_spiral}
\end{equation}
where $\lambda_{x,\text{sp}}$, $\lambda_{y,\text{sp}}$ and $\lambda_{z,\text{sp}}$ being scaling factors carefully selected to produce tight spirals in order to minimise boundary effects, $\gamma=0.5$ for patterns with no variation in contact topology with respect to alternating columns (jersey and garter), $\gamma=0.25$ for patterns with variation in contact topology with alternating columns (rib and seed) and $d$ being the assigned translation distance between rows and it should not exceed the upper bound in order to ensure all rows are attached in initial configuration. At last, we connect all loose ends of yarns through an interpolation scheme to form a complete fabric solely composed of one continuous yarn.

\subsection{Mechanics-centered simulation framework}
\label{sec:2e}
After generating knitted fabrics from previously described geometric model, we performed material characterisation tests to calibrate physical parameters relevant to the simulation summarised in Table \ref{tab1:calibration}. We then apply loading through tethered springs at a sufficiently low loading rate that is critically damped to ensure numerical stability and convergence---see ~\ref{sec:5} for more details.

{\small \begin{center}
\begin{tabularx}{\textwidth}{cXXX}
%\begin{longtable}{|c|c|c|c|}
    %begin{tabular}{llll}
    \hline
    \textbf{Symbol} & \textbf{Parameter} & \textbf{SI in measurements} & \textbf{SI in simulation tests} \\
    \hline
    $\rho$ & mass per unit length & \SI{0.077}{g/m} & \SI{7.7e-4}{g/cm} \\
    $l$ & unit length per stitch & \SI{7.23}{mm} & \SI{0.723}{cm} \\
    $r$ & yarn radius & \SI{524}{\micro \meter} & \SI{0.0524}{cm} \\
    $A$ & yarn cross-sectional area & $A = \pi r^2$ & \SI{8.63e-3}{cm^2} \\
    $I$ & yarn moment of inertia & $I = \pi r^4 / 4$ & \SI{5.92e-6}{cm^4} \\
    $E^s$ & yarn tensile stiffness & \SI{79.0}{MPa} & \SI{7.9e8}{g/cm.s^2} \\
    $E^b$ & yarn bending stiffness & \SI{0.249}{MPa} & \SI{2.49e4}{g/cm.s^2} \\
    $k_g$ & global drag constant & N.A. & \SI{1e6}{g/s} \\
    $k$ & contact repulsive stiffness & N.A. & \SI{1e9}{g/cm.s^2} \\
    $k_{dt}$ & damping constant for tangential frictional force & N.A. & \SI{1e9}{g/cm^2.s} \\
    $k_{dn}$ & damping constant for normal frictional force & N.A. & \SI{1e9}{g/cm^2.s} \\
    \hline
  \caption{Calibration of physical parameters relevant to the simulation of knitted fabrics.}
  \label{tab1:calibration}
  %\end{tabular}
%\end{longtable}
\end{tabularx}
\end{center}}

\vskip6pt

%%%% Section 3 %%%%%%%%%%%%%%%%%%%%%%%%
\section{Results}
\label{sec:results}
%%%% Subsection 3a %%%%%%%%%%%%
\subsection{Effect of pre-tension on validation}
\label{sec:3a}
Experimental evidence shows that despite specifying the same number of stitches along both warp and weft directions, the samples made from same materials but varying stitch patterns often consist of varying dimensions \cite{Allan1983, Amreeva2007, Wei2011}. In addition to slight variation in unit stitch length due to limitations in manufacturing, one key factor for this variability is the internal response of stitch patterns to pre-tension during manufacturing. Typically, yarns are prestressed to be straight and tight when they are fed into the carriers and taken down from the knitting machine. Hence, it is crucial to capture an accurate reference state configuration of knitted fabric under certain pre-tension, in order to provide a meaningful comparison between numerical test and experimental test. Since it is hard to obtain the manufacturing yarn tension a priori, it is hard to establish a relationship between the number of stitches and the fabric tightness analytically. Similar to experimental attempts by Eltahan et al.~\cite{Eltahan2016} and Martinez et al.~\cite{Martinez2021} to find this relationship empirically through regression, we propose to simulate the process of fabrics being prestressed and then relaxed, then calibrate the amount of pre-tension. After the sample is constructed, an initial simulation is performed to reduce the internal rest length of the yarn, so that its configuration becomes tighter---see ~\ref{sec:5d} for more information.

\begin{figure}[htbp]
        \centering \includegraphics[width=0.8\textwidth]{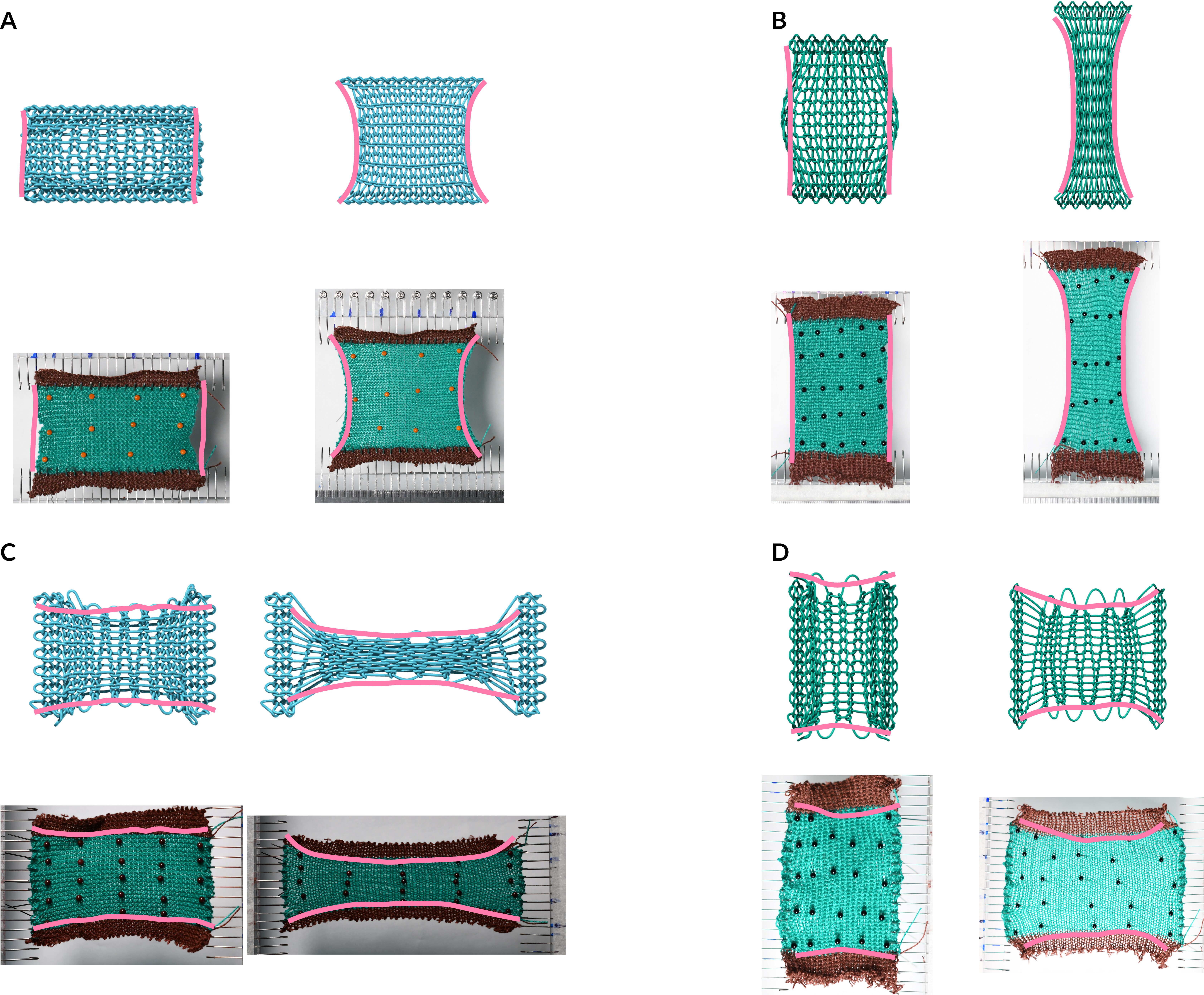}
        \caption{Deformation profiles from simulation (top) and experiment (bottom) at \SI{0}{\%} strain and \SI{80}{\%} strain respectively for (A) garter 1 by 1 when subjected to uniaxial tension along warp direction, (B) rib 1 by 1 when subjected to uniaxial tension along warp direction, (C) garter 1 by 1 when subjected to uniaxial tension along weft direction, (D) rib 1 by 1 when subjected to uniaxial tension along weft direction. Note that free ends of the fabrics with the same pattern and loading condition are marked with the same curves to provide a visual comparison between simulation and experiment on the deformation profiles.}
        \label{fig3a:deform_profile}
    \end{figure}

Though the numbers of stitches used in numerical tests (13 along warp and 12 along weft) are different to those used in experimental tests (41 along warp and 40 along weft), we compare the dimensionless parameters, aspect ratio of fabric and unit arclength of yarn segment, to benchmark numerically obtained configurations in the reference state against manufactured samples. In addition, we calibrate relaxation stage duration by letting the numerical samples to relax in absence of external stress until the side boundaries are curvature-neutral from the untethered boundaries, as shown in Fig.~\ref{fig3a:deform_profile}. By doing so, we were able to obtain close to ground-truth configurations of all four basic weft-knitted fabrics in their reference states.  

%%%% Subsection 3b %%%%%%%%%%%%
\subsection{The fundamental mechanical behaviour of knitted fabrics}
\label{sec:3b}
    \begin{figure}[htbp]
        \centering
        \includegraphics[width=1\textwidth]{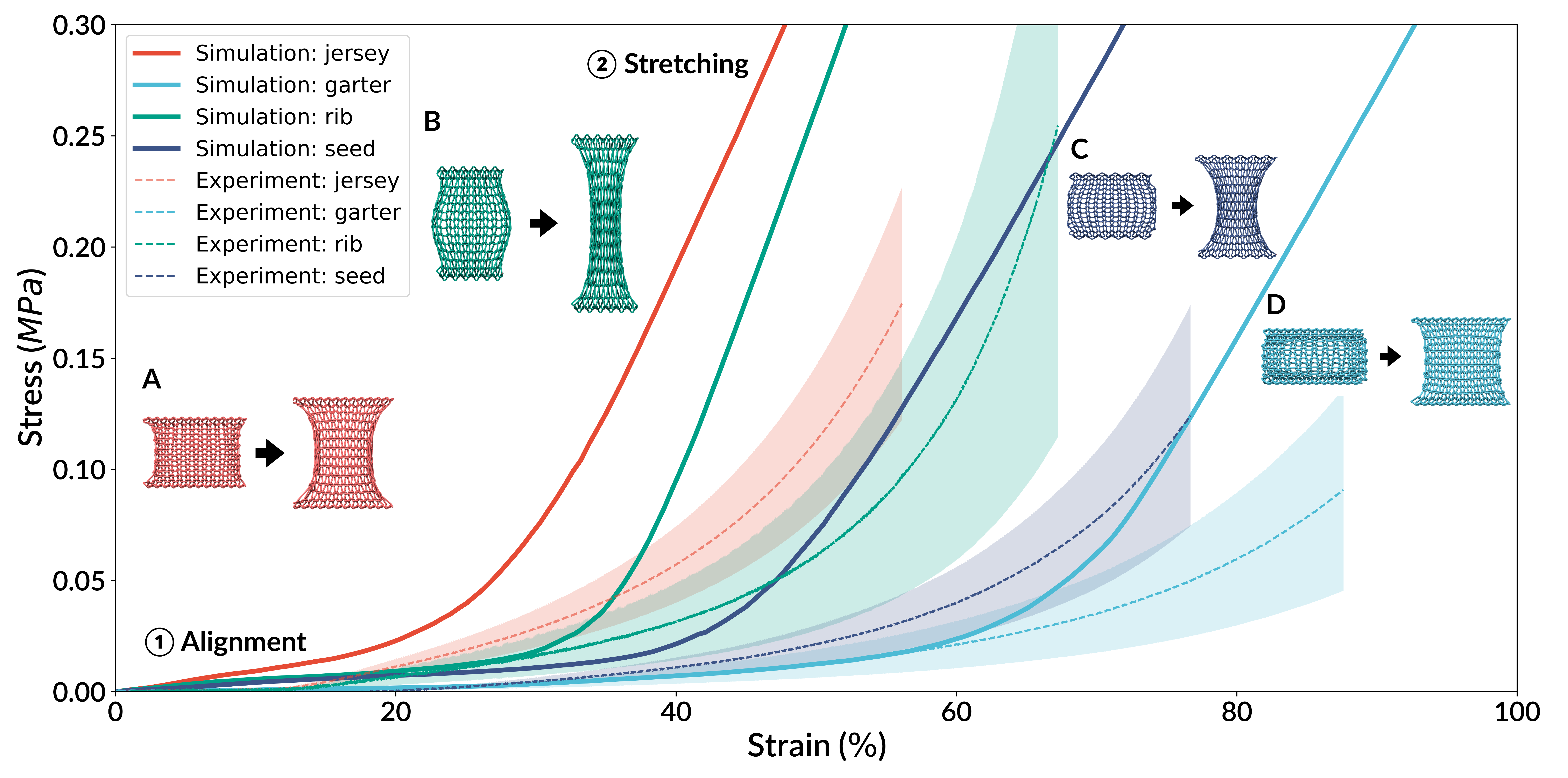}
        \caption{Stress--strain responses measured from simulation (solid lines) and experiment (dashed lines representing averaged responses with shaded area representing standard deviation among all measured samples), and simulation snapshots at \SI{20}{\%} strain and \SI{80}{\%} strain for four weft-knitted fabrics (A) jersey, (B) rib 1 by 1, (C) seed 1 by 1 and (D) garter 1 by 1, when subjected to uniaxial tension along the warp direction.}
        \label{fig3b:stress_strain_warp}
    \end{figure}

    \begin{figure}[htbp]
        \centering
        \includegraphics[width=1\textwidth]{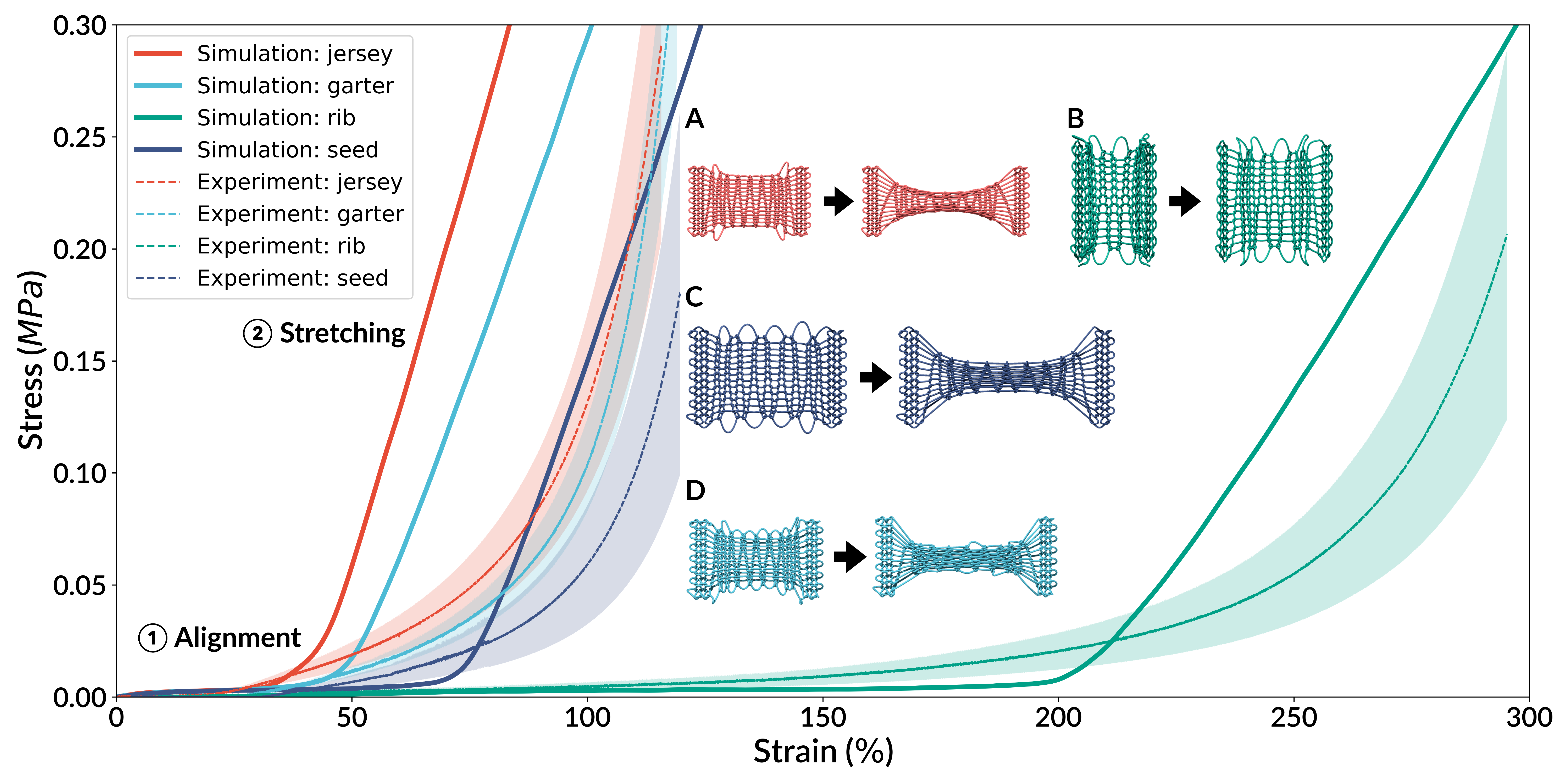}
        \caption{{Stress--strain responses measured from simulation (solid lines) and experiment (dashed lines representing averaged responses with shaded area representing standard deviation among all measured samples), and simulation snapshots at \SI{20}{\%} strain and \SI{80}{\%} strain for four weft-knitted fabrics (A) jersey, (B) rib 1 by 1, (C) seed 1 by 1 and (D) garter 1 by 1, when subjected to uniaxial tension along the weft direction.}}
        \label{fig3b:stress_strain_weft}
    \end{figure}

\noindent The influence of geometric flexibility of fabric pattern on the fundamental mechanical response of knitted fabrics is well captured by our numerical tests and highlighted in Fig.~\ref{fig3b:stress_strain_warp} and Fig.~\ref{fig3b:stress_strain_weft}, which show applied uniaxial tensile stress $\sigma$ against strain $\epsilon$ in warp and weft directions respectively and the corresponding measurements validated by experimental tests. To equalise the difference in the numbers of stitches $n_w$ used in numerically generated fabrics versus those used in machine fabricated samples, we define effective tensile stress as $\sigma = \frac{F}{n_w}$. In addition, to offset variation across varying stitch patterns in fabric dimension $S_R$ along the face formed with yarn diameter $2r$, where the load is applied on, the effective tensile stress is further adjusted to $\sigma = \frac{F}{n_w 2rS_R}$. On the other hand, to equalise fabric dimension $L_R$ along loading direction in the reference state, we define tensile strain $\epsilon = \frac{L_D-L_R}{L_R}\cdot \SI{100}{\%}$ to account for normalised extension $L_D-L_R$. We see good alignment between experiment and simulation, in terms of capturing the overall ``J-shape'' curves and the relative rigidity and extensibility of four knitted fabrics. 

%    \begin{figure}[htbp]
%        \centering \includegraphics[width=0.45\textwidth]{images/fig3b_garter_loady.png}
%        \hskip0.02\textwidth 
%        \includegraphics[width=0.45\textwidth]{images/fig3b_rib_loady.png}
%        \centering \includegraphics[width=0.45\textwidth]{images/fig3b_garter_loadx.png}
%        \hskip0.02\textwidth 
%        \includegraphics[width=0.45\textwidth]{images/fig3b_rib_loadx.png}
%        \caption{Uniaxial tensile tests of two representative samples (a. garter 1 by 1 under tension along warp direction; b. rib 1 by 1 under tension along warp direction; c. garter 1 by 1 under tension along weft direction; d. rib 1 by 1 under tension along weft direction.)}
%        \label{fig3b:rep_samples}
%    \end{figure}
    
Uniaxial tensile tests along the warp direction demonstrate distinct two-stage regions of deformation to failure for all fabrics. Firstly, the jersey fabric, having consistent loop contacts throughout the fabric and hence possessing the simplest geometry, behaved most rigidly among the four fabrics studied during the first stage characterised by low linear stiffness (regime 1) at the magnitude of \SI{0.1}{MPa} up to \SI{30}{\%} strain. Previous experimental studies reported observation of geometry reconfiguration as yarns slide through loop contacts and straighten to align more towards applied tensile load \cite{Sanchez2023}, and our study provides quantitative evidence for such yarn dynamics that are summarised in Sec.~\ref{sec:3c}. After this reconfiguration, the jersey fabric transitioned to a stage where the stiffness monotonically increased by up to \SI{10} times (regime 2), during which stretching of individual yarns become prominent. Statistical measurements are discussed in Sec.~\ref{sec:3c}. Secondly, among the four fabric studied, we observe the garter fabric to be the softest and most stretchable in the initial regime when deformed in the warp direction. This fabric first  undergoes a nearly linear region with the lowest slope (regime 1) at the magnitude of \SI{0.01}{MPa} up to a transitional strain magnitude of \SI{60}{\%}. This regime is followed by yarn stretching (regime 2) at a distinctively higher slope at the magnitude of \SI{1}{MPa}, which continues up to a peak strain approaching \SI{90}{\%}. In contrast, the rib fabric was initially the most rigid and least extensible structure (excluding jersey). This fabric first underwent yarn alignment with an initial slope almost two times higher than that for the garter fabric (regime 1), and with a range only up to \SI{40}{\%} strain, quickly followed by yarn stretching (regime 2) characterised by a much higher slope in the data up to failure at only \SI{60}{\%} strain. In addition, the seed fabric sustained the former loading stage (regime 1) with stiffness similar to that of the rib fabric up to \SI{50}{\%} strain, and transitioned to the latter loading stage (regime 2) with stiffness reaching the asymptotic magnitude of \SI{1}{MPa}.

We observe similar transition behaviour upon uniaxial loading along weft direction, as all fabrics are initially soft and stretchable, followed by strain hardening as geometric flexibility from the mesoscale patterns are exhausted. However, the relative rigidity and extensibility of fabrics are now different. Though the jersey fabric is again the most rigid and undertook strain-hardening the soonest at \SI{30}{\%} strain during loading regime 1, the relative variation in stiffness among the four fabrics during this loading regime is negligible. The rib fabric, previously representative of rigid behaviour under tension along warp direction, now becomes the initially softest and most stretchable under tension along weft direction, as it first undergoes yarn alignment of a (regime 1) at almost negligible magnitude up to more than \SI{100}{\%} strain, followed by yarn stretching (regime 2) at a noticeably higher slope at the magnitude of \SI{0.1}{MPa} in the data up to failure strain more than \SI{150}{\%}. Conversely, the garter fabric under tension along weft direction is the most rigid and least extensible in the initial regime (excluding jersey), as it first underwent yarn alignment with a slope almost two times higher than that for the rib fabric (regime 1), and only had a comparably small range up to \SI{50}{\%} strain. This regime is succeeded by yarn stretching (regime 2), exhibiting a markedly higher slope nearing \SI{1}{MPa} and leading to failure at nearly \SI{100}{\%} strain. In addition, the seed fabric undertook mechanical behaviour close to that of the garter fabric.

%\KB{here I suggest to comment on the non-perfect agreement between exp and simulations? What are the contributing factors? Can we also compare experimental and numerical deformed shapes of the knits?}

It is well accepted that the precise matching between reduced-order constitutive models and experimental tests on knitted fabric is challenging \cite{Vassiliadis2007, Yeoman2010, Dinh2018, Liu2017, Weeger2018, Wu2020}. Though our model recovers the general two-stage nonlinear elastic behaviour and the relative responses among the four basic weft-knitted fabrics well, we notice the limitations, particularly when capturing the behaviour when fabrics transition between regimes. There are multiple reasons for these discrepancies, such as our treatment of the yarn as a solid elastic tube, which may not precisely capture the spun fibers in the acrylic yarn. In addition, we assume a linear stretching force response in the yarn, which may not be accurate at high strains, as acrylic yarn typically softens at high strains as characterised in ~\ref{sec:6d}. We further discuss this limitation in Sec.~\ref{sec:3c}, where we collect statistical measurements of individual yarn segment stretch for all studied fabrics and loading conditions.  

%%%% Subsection 3c %%%%%%%%%%%%
\subsection{Mechanical role of yarn dynamics on fabric extensibility and anisotropy}
\label{sec:3c}
\begin{figure}[htbp!]
        \centering \includegraphics[width=0.6\textwidth]{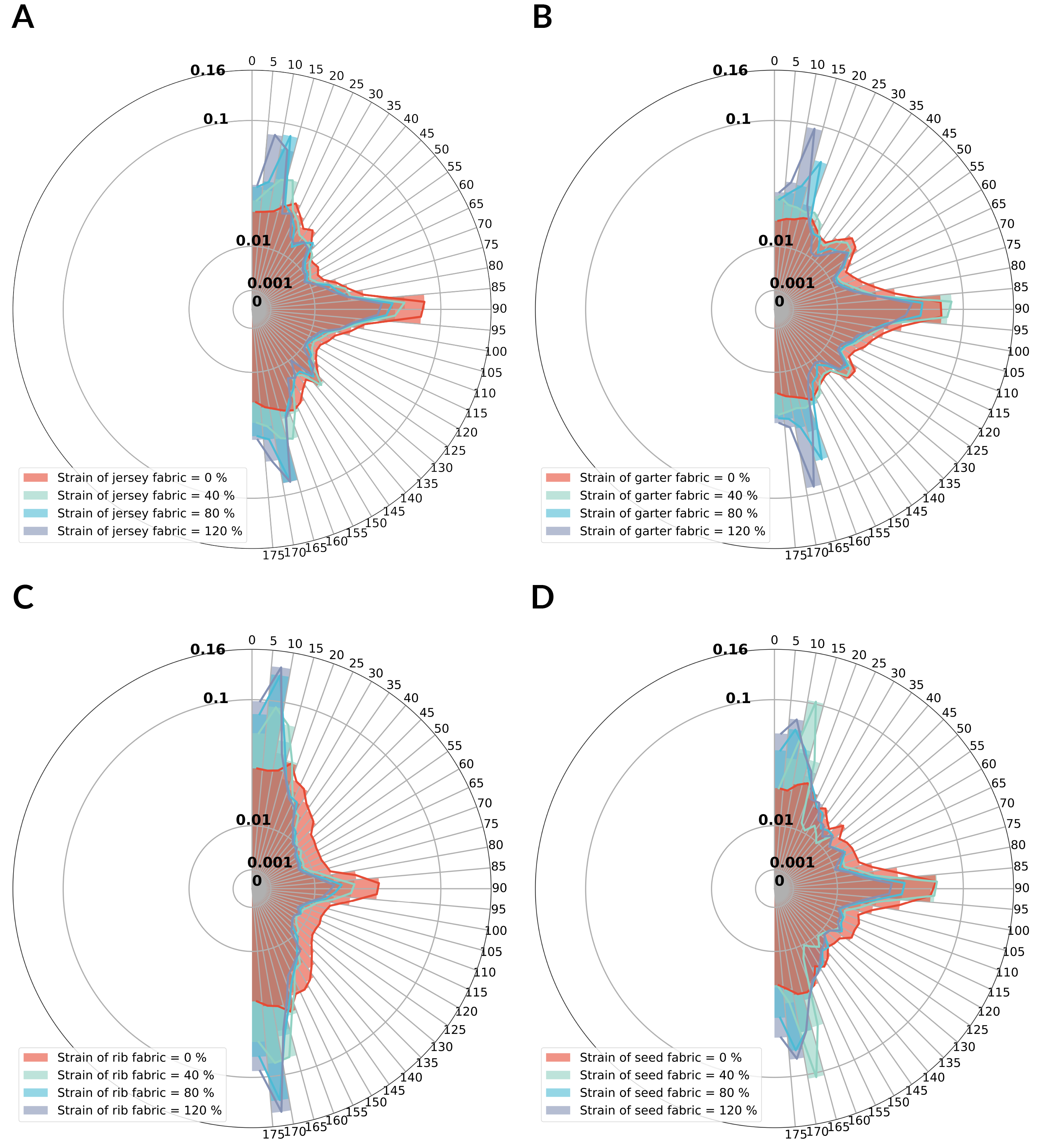}
        \caption{{Rose diagram measuring the alignment of yarn segments with respect to the loading direction for all samples: (A) jersey, (B) garter 1 by 1, (C) rib 1 by 1 and (D) seed 1 by 1, under uniaxial tension along warp direction.}}
        \label{fig3c:rose_loady}
    \end{figure}

\begin{figure}[htbp!]
        \centering \includegraphics[width=0.6\textwidth]{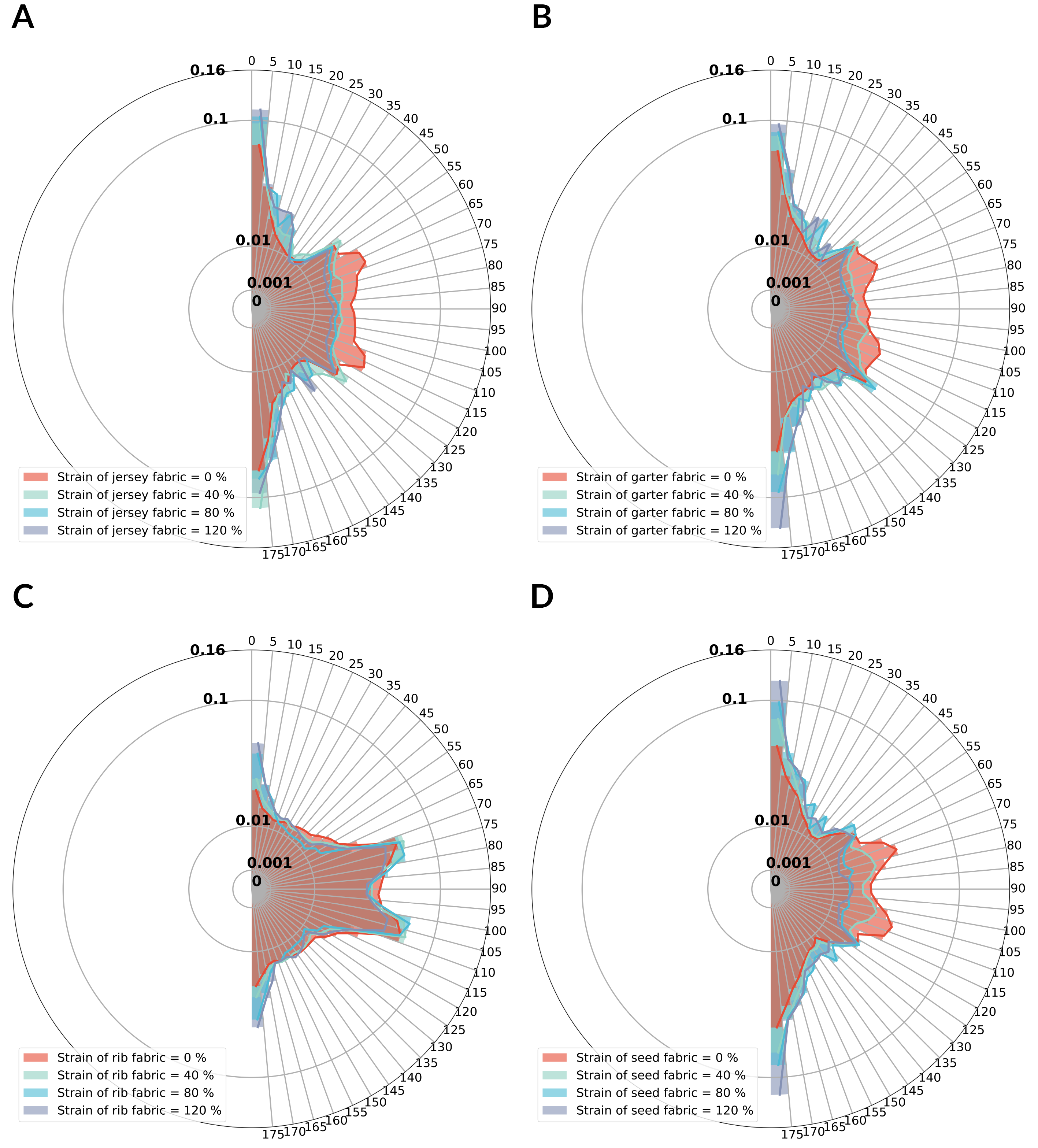}
        \caption{{Rose diagram measuring the alignment of yarn segments with respect to the loading direction for all samples: (A) jersey, (B) garter 1 by 1, (C) rib 1 by 1 and (D) seed 1 by 1, under uniaxial tension along weft direction.}}
        \label{fig3c:rose_loadx}
    \end{figure}

\begin{figure}[htbp!]
        \centering \includegraphics[width=0.6\textwidth]{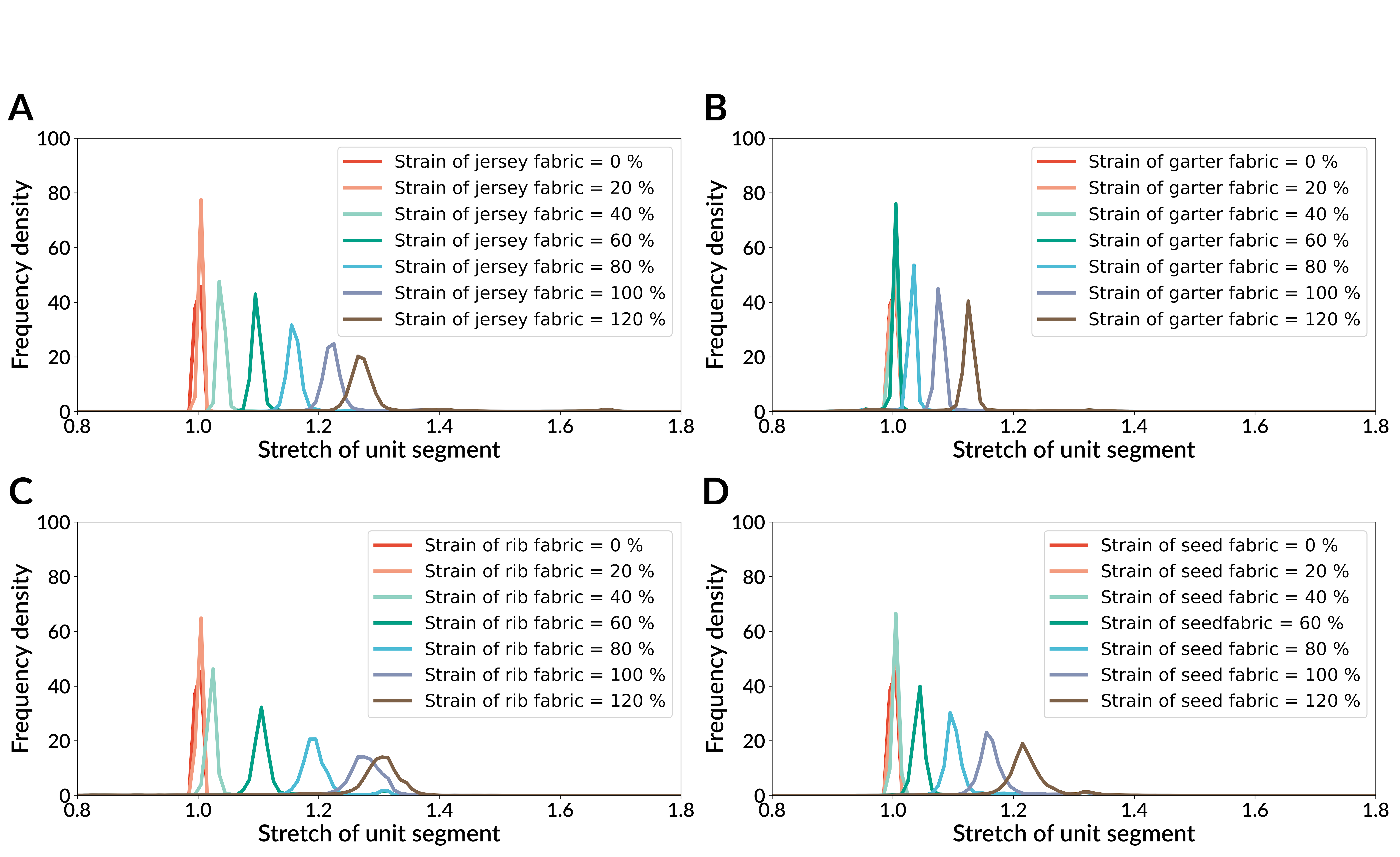}
        \caption{Histogram measuring the distributions of yarn segment stretch for all samples: (A) jersey, (B) garter 1 by 1, (C) rib 1 by 1 and (D) seed 1 by 1, under uniaxial tension along warp direction.}
        \label{fig3c:hist_loady}
    \end{figure}

\begin{figure}[htbp!]
        \centering \includegraphics[width=0.6\textwidth]{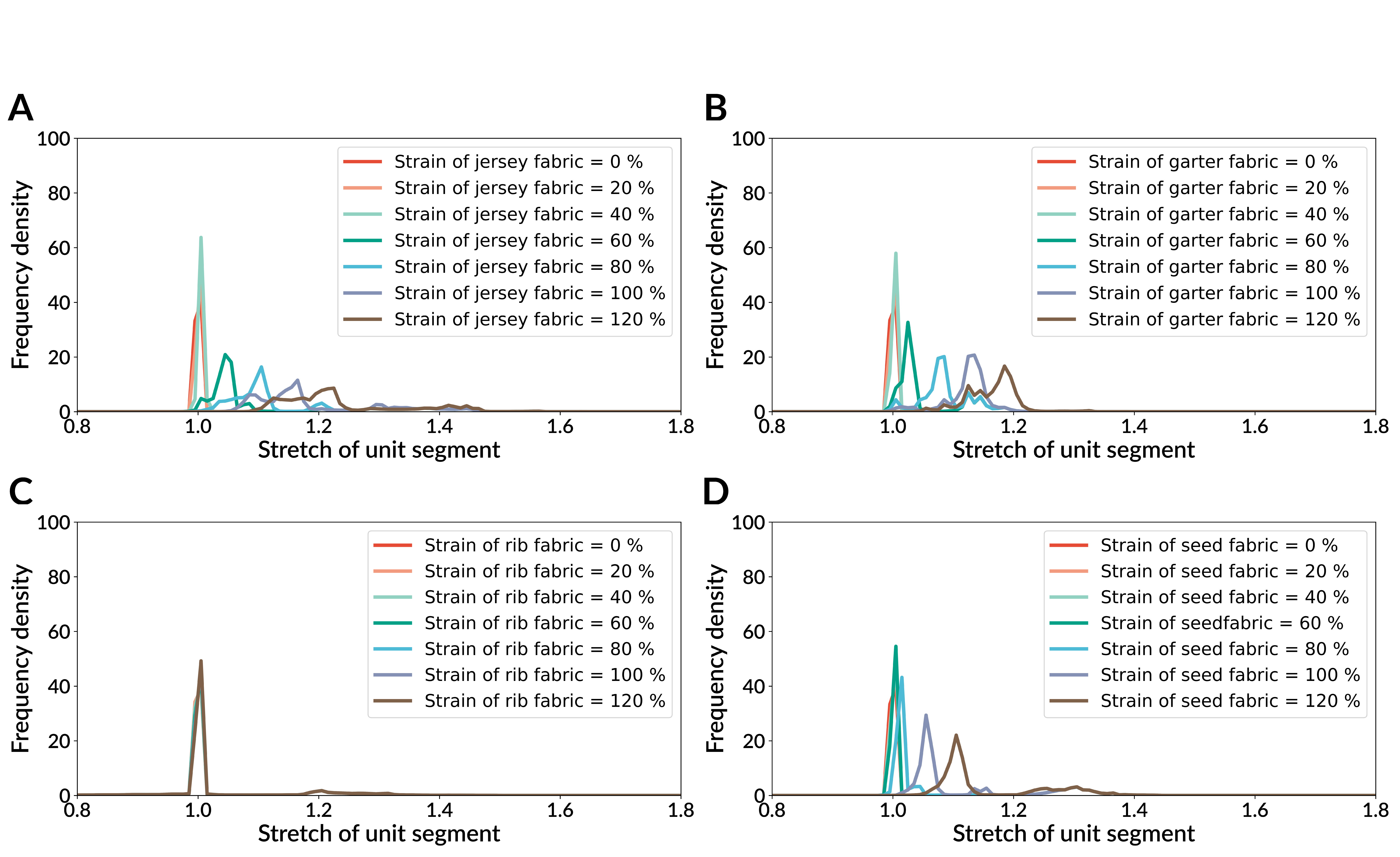}
        \caption{Histogram measuring the distributions of yarn segment stretch for all samples: (A) jersey, (B) garter 1 by 1, (C) rib 1 by 1 and (D) seed 1 by 1, under uniaxial tension along weft direction.}
        \label{fig3c:hist_loadx}
    \end{figure}

To probe into how yarn rearrangements influence the macroscopic extensibility of knitted fabrics, we measured the projection of individual yarn segments on to the loading direction of warp in Fig.~\ref{fig3c:rose_loady} and of weft in Fig.~ \ref{fig3c:rose_loadx}. With an angular increment (bin size) of \SI{5} degrees, we tracked the overall evolution of yarn segments as they aligned closer with the applied load as tensile strain increased from \SI{0}{\%} to \SI{120}{\%}. This quantitative evidence compliments experimental observation of reorientation of yarn segments to exploit geometric degrees of freedom within the connected network. Such geometric rearrangement of yarn segments rather than material stretching of yarn segments contributes to the compliant behaviour during the initial stage of fabric mechanical response. Statistical distributions of yarn segment stretch in Figs.~\ref{fig3c:hist_loady} \& \ref{fig3c:hist_loadx} further support the yarn reorientation mechanism during the initial loading on fabrics, as distribution peaks remain within the range for negligible segment stretch while fabrics are stretched until transitions occur. It is worth noting that even as fabrics transition to higher strain ranges (near or exceeding \SI{100}{\%}) and the most stretched segments approach \SI{20}{\%}, these segments account for less than \SI{20}{\%} of all yarn segments. Therefore, it is reasonable to assume linear elasticity for the majority of yarn segments as an averaged one-time calibration of the stretching stiffness for preliminary study. However, this places challenges in addressing fabric behaviour between regimes, as inhomogeneous mechanical field of segment stretch contributes to the transition. 

Previously, we also observed anisotropy from knitted fabrics, which is topology-dependent. Considering garter and rib as representative examples, we establish the former has a softer mechanical response and sustains a higher elastic strain range when subjected to tension along fabric warp direction, while the latter behaves in a stiffer manner within a lower elastic strain range when responding to tension along fabric weft direction. To probe into the influence of fabric structure on anisotropy, we begin by examining their geometric configurations in the reference state, considering yarn segment angles in Figs.~\ref{fig3c:rose_loady} \& \ref{fig3c:rose_loadx}. The garter fabric initially has less than \SI{10}{\%} of yarn segments aligning with the warp direction within a difference of \SI{10} degrees, but has more than \SI{20}{\%} similarly close yarn alignment with the weft direction. In comparison, the rib fabric has more than \SI{20}{\%} of yarn segments closely aligned with  the warp direction in the reference state, and only less than \SI{15}{\%} yarn segments aligned with the weft direction at this stage. The lower initial frequency of alignment with the loading direction provides more geometric degrees of freedom for the yarn segments reorient themselves to manifest applied stress within the hierarchical system, making the fabric more compliant under this ``unaligned'' loading direction than the orthogonal direction. Moreover, as observed in Figs.~\ref{fig3c:hist_loady} \& \ref{fig3c:hist_loadx}, for a fabric to demonstrate softer and more compliant behavior under a fixed loading condition, the shift of its peak yarn segment stretch distribution towards a higher stretch range occurs more slowly. 

%%%% Subsection 3d %%%%%%%%%%%%
\newpage
\subsection{Demonstration of the design space}
\label{sec:3d}
%\KB{this seems to be way too qualitative for my taste. You should describe what you are showing in the Fig. Which combinations are you considering? How did use your previous results to choose these combinations? Is the behavior as expected?} 

\begin{figure}[!htbp]
        \centering \includegraphics[width=0.8\textwidth]{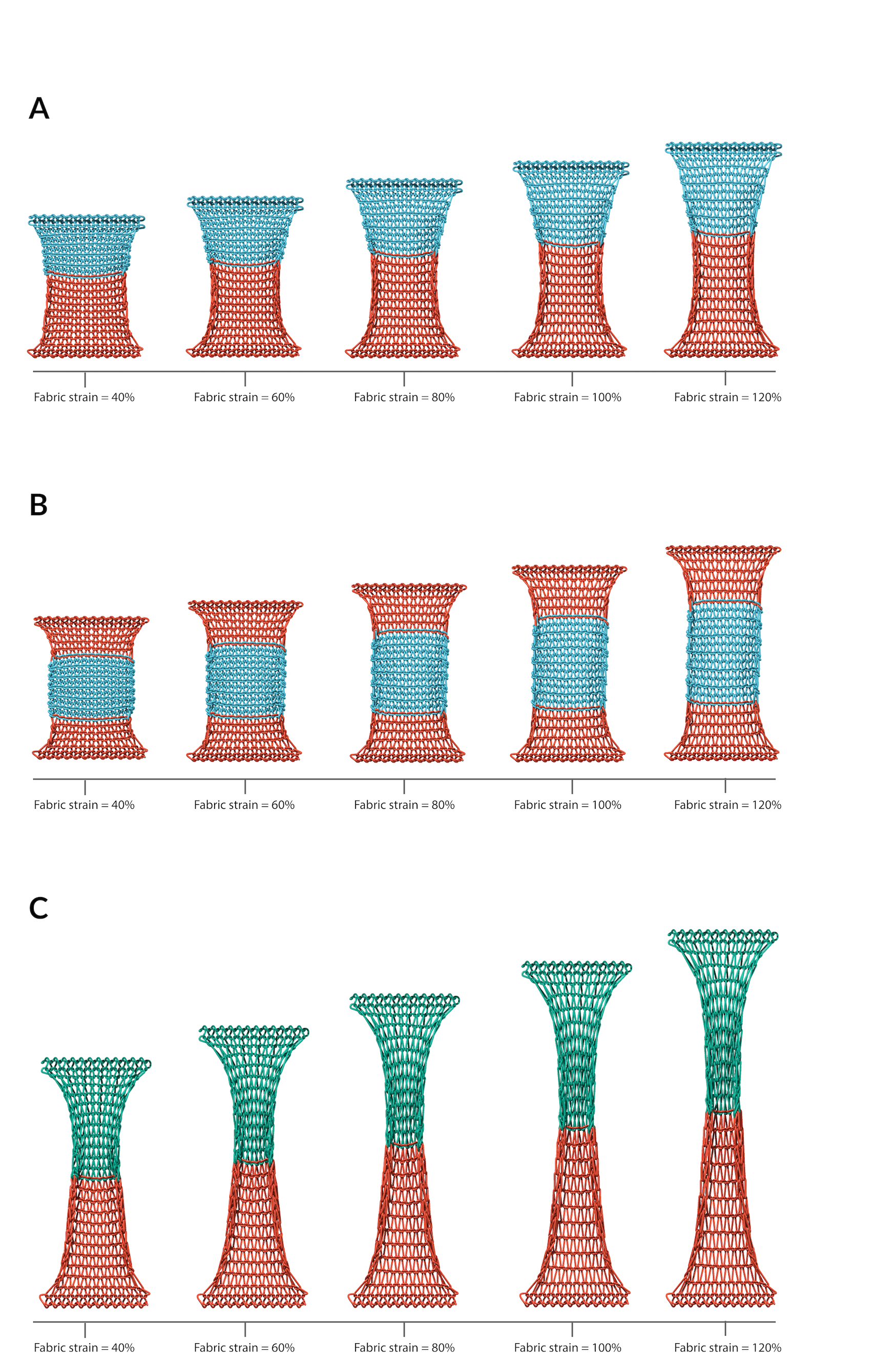}
        \caption{Gallery of knitted fabrics consisting of varying topology and spatial distributions (A) top half: garter 1 by 1, bottom half: jersey; (B) top quarter: jersey, middle half: garter 1 by 1, bottom quarter: jersey; (C) top half: garter 1 by 1, bottom half: jersey stretched with uniaxial tension along warp direction ranging from \SI{40}{\%} to \SI{120}{\%} strain. Note that varying knit structures are colour-coded here.}
        \label{fig3d:multi_patterns}
    \end{figure}

Guided by our study on the anisotropy of basic weft-knitted structures in Sec.~\ref{sec:3b}, we can purposefully explore the design space for textile-based devices to deform to desired shapes. First, we highlight how structural variation in multi-structure knitted fabrics consisting of the same numbers of stitches along fabric warp and fabric weft leads to a wide range of compliance, when they are subject to the same tensile load in Fig.~\ref{fig3d:multi_patterns}. 
Here, both asymmetric primitive (A) and symmetric primitive (B) utilise the relatively better stretchability of garter over jersey along the loading direction (fabric warp), leaving more fabric to be distributed along the orthogonal direction that can provide localised comfortability to the wearer along fabric weft. The former provides looser fit at the fabric boundary, while the latter provides looser fit at fabric middle region. 
On the other hand, asymmetric primitive (C) shows how using a less stretchable structure (rib) along the loading direction (fabric warp) potentially enhances gripping capability of a device upon actuation, as fabric quickly curls out of plane along the direction orthogonal to actuation and forms pocketed region. This can be directly applied as responsive structures to be passively actuated and textiles to provide custom fit.

\begin{figure}[!htbp]
        \centering \includegraphics[width=0.7\textwidth]{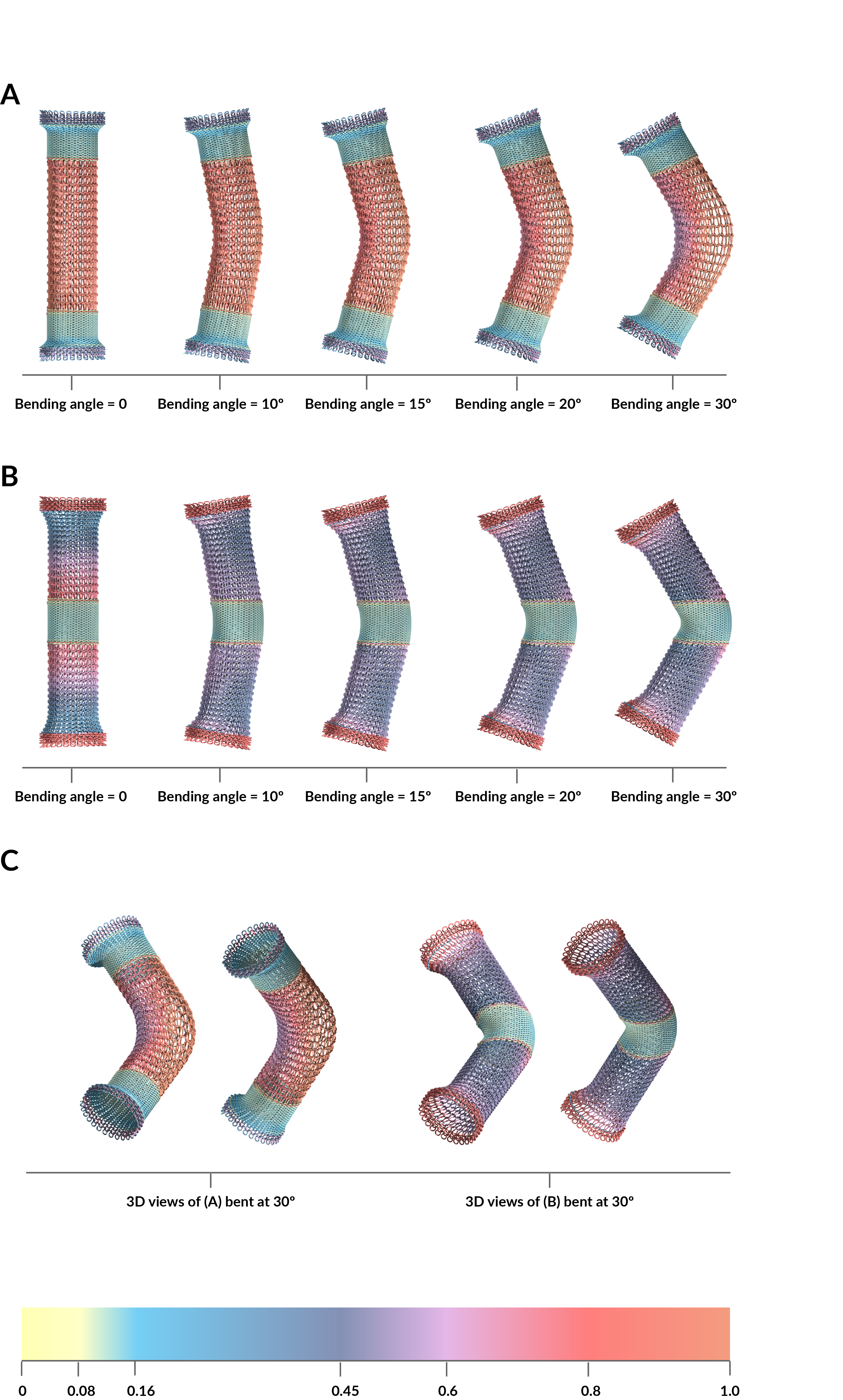}
        \caption{{Demonstration of actuated 3D composite tubes made of jersey knits (A) 2D projection to show material distribution: two ends with rigid materials, middle with soft materials bent from \SI{0}{} to \SI{30}{degrees}; (B) 2D projection to show material distribution: two ends with soft materials, middle with rigid materials bent from \SI{0}{} to \SI{30}{degrees}; (C) 3D views of actuated 3D composite tubes. Note that the colour variation is based on stretching energy per unit arc length, of which the comparative scale from \SI{0}{} to \SI{0.08}{J/cm} is shown in attached colour map.}}
        \label{fig3d:composite_patterns}
    \end{figure}

In addition, we demonstrate the adaptability of our model to vary material properties at the yarn level and the generalisability of our model to create 3D configurations, both further opening up the design space of functional textiles to composites. Fig.~\ref{fig3d:composite_patterns} shows the deformation processes of two 3D primitives made of jersey throughout with (A) having more rigid materials at both ends and (B) having softer materials at both ends, both subject to bending applied through compressed tethered boundaries. This is a demonstration of direct application in soft actuators to absorb impact. Though failure is not included in the scope of this work, we apply a custom colour map in Fig.~\ref{fig3d:composite_patterns} based on stretching energy of individual yarn segments, to highlight the capability of our model to investigate micromechanical hot spots due to inhomogeneity inherent across the whole fabric. As expected, we see the outer side stretches more than the inner side upon bending. Moreover, we see that the region consisting of more rigid material stretches less than that consisting of softer material.

\vskip6pt

%%%% Section 4 %%%%%%%%%%%%%%%%%%%%%%%%
\section{Conclusion}
\label{sec:conclusion}
%%%% Subsection 4 %%%%%%%%%%%%
\label{sec:4}
In summary, we first present the mechanics of knitted fabrics through micromechanical lenses from yarn dynamics. By defining a dynamic formulation of the governing equation, with simple yet adaptive constitutive law at each yarn segment, we have developed and implemented a computational model to efficiently solve for the evolution of such complex system with localised mechanical fields. Our numerical study complimented by experimental evidence show the fundamental mechanical response of knitted fabrics, noticeably characterised by J-shape behaviour analogous to hierarchical biological structures with geometric degrees of freedom arising from the separation of scales \cite{Meyers2013, Barthelat2016, Oosten2019, Zhang2021}, echoing the current wider interest in understanding how soft materials gradually adapt to applied elastic energy. In particular, we include the effect of pre-tension in our numerical procedure, in order to provide meaningful comparison with experimental measurement. In addition, we probe into the topology-dependent variation in fabric stiffness, extensibility and anisotropy by conducting parametric study on a set of representative weft-knitted fabrics. Supported by statistical measurements of inhomogeneous yarn segment stretch and alignment that are not feasible from experiments, we provide insights on the remarkable differences among weft-knitted fabrics from varying topological groups. Last but not least, we demonstrate how to apply learnt mechanical properties of varying stitch patterns to manipulate the design for targeted responses and localised compliance. Such multi-structure multi-material configurations in both 2D and 3D, of which the enormous design space can be explored by rapid deployment of our computational tool. By doing so, we hope to pave the path for systematic design of mechanically programmable fabrics and textiles beyond what their constitutive materials can achieve through demonstrations in responsive structures, wearables and soft actuators.
\vskip6pt

%%%% Section 5 %%%%%%%%%%%%%%%%%%%%%%%%
%\section{Methods}
\appendix
\label{sec:methods}
%%%% Section 5 %%%%%%%%%%%%
\section{Implementation details of the numerical method}
\label{sec:5}
\subsection{Numerical integration and performance}
\label{sec:5a}
The numerical simulations in this paper are performed using a custom C++ code that uses the OpenMP library~\cite{dagum98} for multithreading. The core of the
simulation involves integrating the ODE system for the position $\vq_k$ and velocity $\vqd_k$ degrees of freedom described in Eqs.~\eqref{eq:ys} \&
\eqref{eq:vs}. This is solved using the fourth-order adaptive ``first same as last'' (FSAL) Runge--Kutta method~\cite{hairer93}. This method uses five
intermediate stages, where the first four can be used to construct a fourth-order accurate solution, and the final stage can be used to construct an auxiliary third-order accurate solution for step size selection. With the FSAL property, the final stage can be re-used as the first stage of the next step, reducing the total computational work. We implement adaptive step size selection via the
procedures described by Hairer et al.~\cite{hairer93}, which use a combination of absolute tolerance $\Atol$ and relative tolerance $\Rtol$. Initial step size
selection is also performed using the methods of Hairer et al.~\cite{hairer93}.

Adaptive integration is beneficial for our simulations, since the appropriate step size varies greatly over the course of the simulation. In the initial stages, yarn elasticity is the most important physical effect and large
timesteps can be taken. Once contact forces become important, the timestep sizes are substantially reduced. Similar methods have been employed in the simulation of crumpled sheets~\cite{andrejevic22}, which have comparable
behavior.

Our simulations output snapshots at equally spaced time intervals. Since the timesteps are chosen adaptively, the integration time points will generally not align with the output time points. To solve this issue, we make use of
dense output~\cite{hairer93} and construct a cubic Hermite interpolant of the simulation state over each integration time step. Evaluating this interpolant at the output time point results in a fourth-order accurate approximation of
the solution. The snapshots are outputted as binary files that contain the complete simulation state, which can be post-processed to perform a variety of analyses.

It is worth noting that the ODE system representing the yarn mechanics is not infinitely differentiable, since the contact forces are discretely switched on and off as yarns move past each other. Proving that the Runge--Kutta scheme is fourth-order accurate requires that the mathematical solution has Taylor expansions up to fourth order, which is not true in the case when an integration timestep passes over a moment when a contact force is switched on
or off. Because of this scenario, it is not possible to guarantee that the results are fourth-order accurate. Nevertheless, we opt to use the fourth-order scheme
since it results in good accuracy and performance overall. Furthermore, Hairer et al.~\cite{hairer93} demonstrate that in practical cases, adaptive-timestep integrators can approach high-order accuracy even when the ODE system lacks
sufficient smoothness, since the integrator can automatically refine the timestep when passing over a discrete switch in the ODE, minimizing the additional error incurred.

{While our paper focuses on knitted samples comprising of a single long yarn, our code can general configurations with multiple yarns. Supplemental Information (SI) Fig.~1 demonstrates the capability of the code to handle braided structures and woven fabrics. A theoretical analysis of the algorithms shows that the simulations scale linearly with the total yarn segments. In SI Fig.~2 we confirm this behavior for both knitted fabrics and woven fabrics, and also demonstrate good parallel scaling with multithreading.}

\subsection{Linear system}
\label{sec:5b}
For a single yarn with $N$ spline segments we write $\vq\in \R^{3(N+3)}$ and $\vqd\in \R^{3(N+3)}$ to be the vectors describing the yarn position and
velocity, respectively. From Eqs.~\eqref{eq:lagr} \& \eqref{eq:dkint} the
general equation of motion for a particular component $(\vq_k,\vqd_k)$
is given by
\begin{align}
  \frac{d}{dt} \left( [ M \vqd]_k \right) &= \Big( -\nabla_{\vq_k} V(\vq) - \nabla_{\vqd_k} D(\vq,\vqd) \Big), \label{eq:lagr2a} \\
  \frac{d}{dt} \left( \vq_k \right) &= \vqd_k \label{eq:lagr2b}
\end{align}
where $M$ is a banded matrix whose components are defined by
Eq.~\eqref{eq:ke_integ}. For $2\le j \le N-2$, away from the end points, the
components of $M$ are given by
\begin{equation}
  M_{jk} = \begin{cases}
    \tfrac{151}{315} & \qquad \text{if $k=j$,} \\
    \tfrac{397}{1680} & \qquad \text{if $|k-j|=1$,} \\
    \tfrac{1}{42} & \qquad \text{if $|k-j|=2$,} \\
    \tfrac{1}{5040} & \qquad \text{if $|k-j|=3$,} \\
    0 & \qquad \text{otherwise.}
  \end{cases}
\end{equation}
Near the end points, the matrix values change, because the B-spline
functions are no longer fully contained within $\Omega$. The values for $j<2$
are given in Table \ref{tab:ke_boundary}, and the values for $j>N-2$ are
obtained via symmetry. Our code can also handle the case when either the
position or direction of the end point is fixed, which results in adjusting
the linear system to incorporate a linear algebraic constraint.

\begin{table}[!ht]
  \centering
  \begin{tabular}{|c|cccccc|}
    \hline
    & $k=-1$ & $k=0$ & $k=1$ & $k=2$ & $k=3$ & $k=4$ \\
    \hline
    $j=-1$ & \nicefrac{1}{252} & \nicefrac{43}{1680} & \nicefrac{1}{84} & \nicefrac{1}{5040} & & \\
    $j=0$ & \nicefrac{43}{1680} & \nicefrac{151}{630} & \nicefrac{59}{280} & \nicefrac{1}{42} & \nicefrac{1}{5040} & \\
    $j=1$ & \nicefrac{1}{84} & \nicefrac{59}{280} & \nicefrac{599}{1260} & \nicefrac{397}{1680} & \nicefrac{1}{42} & \nicefrac{1}{5040} \\
    \hline
  \end{tabular}
  \caption{Coefficients in the matrix $M$, defined in Eq.~\eqref{eq:ke_integ}
  arising from the kinetic energy term in the Lagrangian formulation
  of the yarn dynamics.\label{tab:ke_boundary}} 
\end{table}

To solve the ODE system in Eqs.~\eqref{eq:lagr2a} \& \eqref{eq:lagr2b} it is
necessary to solve the linear system $Mq=f$ where $q\in \R^{(N+3)\times 3}$
are entries of $\vqd$ arranged into a matrix, with the $x$, $y$, and $z$
components each in one column. $f \in \R^{(N+3)\times 3}$ are the
corresponding source terms, from the right hand side of Eq.~\eqref{eq:lagr2a}. The
matrix $M$ remains fixed throughout the simulation, and therefore during the
initialization its LU factorization is precomputed. This accelerates the
solution of the linear system during the simulation. The LAPACK library is
used, with the LU factorization being performed using the \texttt{dgbtrf}
routine for a general banded matrix in double-precision floating point
arithmetic. The \texttt{dgbtrs} routine is then used to solve the linear
systems during the timesteps.

%\begin{table}
%  \begin{tabular}{ccccccc}
%    ,1/6.,2/3.,1/6.,0.,
%    ,7/720.,173/1260.,137/840.,29/1260.,1/5040.,
%    ,121/1260.,13093/5040.,4553/630.,9883/2520.,509/1260.,17/5040.,
%  \end{tabular}
%
%  \begin{tabular}{ccccccc}
%    ,1/2.,0.,-1/2.,0.,
%    ,43/1680.,151/630.,59/280.,1/42.,1/5040.,
%    ,1/63.,397/1680.,307/630.,149/630.,1/42.,1/5040.,
%  \end{tabular}
%
%  \begin{tabular}{cccccc}
%    ,1/6.,2/3.,1/6.,0.,
%    ,1/2.,0.,-1/2.,0.,0.,
%    ,31/5040.,587/2520.,55/72.,283/630.,239/5040.,1/2520
%  \end{tabular}
%\end{table}

\subsection{Contact sphere diameter calculations}
\label{sec:csphere}
As described in Sec.~\ref{sec:2c}, yarn--yarn contact forces are handled by
introducing $n$ equally spaced spheres of radius $d_\text{con}$ along each spline segment.
If $\dcon=d$, then the envelope formed by the spheres would be smaller
than the yarn itself. We therefore systematically choose $\dcon$ to
better approximate the yarn shape. Let $D=l/(2n)$ be the distance between
successive contact spheres. Assuming that $D$ is small relative to $d$, the
contact sphere diameter is chosen to be
\begin{equation}
  \dcon=\frac{d+\sqrt{d^2 + \tfrac23 D^2}}{2},
\end{equation}
which ensures that the average diameter of the envelope of spheres is equal to
$d$. To choose the number of contact spheres, we define a parameter $\alpha$
corresponding to the maximum allowable mean square deviation between the
contact sphere envelope and the filament diameter, which is typically set to be
several percent. Then the number of contact spheres satisfies
\begin{equation}
  n = \left\lceil \frac{l}{d\sqrt{6(1+\alpha)\alpha}}  \right\rceil,
\end{equation}
where $\lceil\cdot\rceil$ is the ceiling operator. Since the contact spheres
overlap, when considering Eq.~\eqref{eq:contact_integ}, it is necessary to
screen out the effect of interactions from neighboring spheres along the same
yarn. This is done by defining a screening number $n_\text{screen} = \lceil
\beta \dcon/D \rceil$ where $\beta$ is a dimensionless parameter. Terms in the
sum of Eq.~\eqref{eq:contact_integ} are only considered if the indices of the
spheres are separated by at least $n_\text{screen}$.

\subsection{Tethering forces and initial sample generation}
\label{sec:5d}
To perform the uniaxial tension tests, tethering forces are applied to the
fabric to fix the displacement in two end regions. This procedure is similar
to the approach used in the immersed boundary method~\cite{peskin02}
to simulate fixed walls~\cite{fai18}. Specifically, two regions $D_+$ and
$D_-$ are defined, where typically $D_\pm = \{ (x,y,z)\in \R^3 \,:\, \pm
y>y_\text{fix} \}$ for a constant $y_\text{fix}$ when pulling a sample in the
$y$ direction.

During the simulation initialization, all spline segments that lie fully within
$D_+$ and $D_-$ are marked, and the reference position of each
quadrature point within each marked segment is recorded. Using
this information, the additional energy contributions
\begin{equation}
  V^t_\pm = k_t \int_\Omega I_{D_\pm}(s) \| \vy(s) - \vec{F}_\pm(\vy_\text{ref}(s),t) \|^2 ds
  \label{eq:energy_tether}
\end{equation}
are added to Eq.~\eqref{eq:lagr}, where $I_{D_\pm}(s)$ is equal to one in spline
segments marked within $D_\pm$, and zero otherwise. Here, $\vec{F}_\pm: \R^3
\to \R^3$ are time-dependent affine transformations of the reference position.
They can be used to apply the constant pulling velocity in the end regions.
The forces that are measured in the tension tests are computed as the total
force applied to the fabric in each tethered region.

As described in ~\ref{sec:6e}, the experimental samples created by the knitting
machine are already under substantial tension, meaning that the yarns are much
tighter than the examples shown in Figs.~\ref{fig2:knit_patterns} \&
\ref{fig1:knit_assembly}. It is difficult to initialize the simulation in tighter
configurations directly, since any overlaps in the initial state may result in
very large initial contact forces. Because of this challenge, we perform a preliminary
simulation to generate the samples for the tension tests. We initialize the
yarn in a loose configuration given by Eq.~\eqref{eq:basic_param_spiral}, and
then make the spline rest length $l$ a function of time, applying a linear
ramping so that
\begin{equation}
  l(t) = \begin{cases}
    l_0 - (1-\eta) \frac{t}{T_r} & \qquad \text{if $t<T_r$,} \\
    \eta l_0 & \qquad \text{if $t\ge T_r$,}
  \end{cases}
\end{equation}
where $l_0$ is the initial rest length of the yarn, $T_r$ is the duration of
the ramping, and $\eta$ is a dimensionless ratio chosen to ensure that the
final state matches the same compression as the experimental samples. During
this procedure, the two end regions are tethered using the energy contributions
in Eq.~\eqref{eq:energy_tether} to prevent the sample from curling. Since the
rest length is reduced, the affine transformations $\vec{F}_\pm$ are used to
apply a commensurable shrinkage to the tethered regions. The precise amount of
shrinkage is determined by comparing to the geometry of the
experimental samples. After this preliminary simulation is performed, the yarn
state is saved and then used as the starting configuration for the tension test
simulations.

\section{Materials and Experiments}
\label{sec:6}
\subsection{Material selection}
\label{sec:6a}
Commercially available acrylic spun yarns (16/2 Vybralite Acryclic Yarns, National Spinning Co. obtained at Peter Patchis Yarns, USA) were selected to create all knitted fabric swatches. We performed the tensile tests  and bending tests on single yarns, in order to benchmark the previously reported stretching stiffness, and to calibrate the bending stiffness of the selected experimental material.
  
\subsection{Filament diameter}
\label{sec:6b}
Average effective cross-sectional diameters of the yarn, ``filament diameter,'' were obtained by imaging \SI{20}{mm} lengths of yarn at rest using a laser microscope (Olympus OLS4000). We utilized Adobe Photoshop to compute the projected area of the yarn segment under orthogonal projection, which we then divided  by the fixed yarn length of \SI{20}{mm} to determine the filament diameter. We repeated this process with three different yarn segments from the same material to ensure accuracy. The averaged yarn radius, \SI{0.0524}{cm}, from experimental measurements was used to benchmark the yarn diameter used in simulation. In addition, we assume a circular and consistent cross section with this calibrated effective radius from onward for related calculations. 

\subsection{Yarn linear density}
\label{sec:6c}
We adopt yarn linear density defined as 
\begin{equation}
  \rho = \frac{M}{l},
  \label{eq:yarn_density}
\end{equation}
which is a standard definition adopted in the textile industry and was characterized from measuring samples as \SI{0.077}{g/m}. The equivalent term in simulation is \SI{7.7e-4}{g/cm}. Simulating this density directly is challenging due to very fast propogation of elastic modes that requires small timesteps to resolve. We therefore boost this density be a scaling factor of $10^9$, which effectively lowers the elastic wave speed by $10^{4.5}$. This change only affects dynamical behavior of our simulations and does not strongly alter the measured stress-strain curves that are in a quasi-static regime.

\subsection{Characterisation of yarn properties}
\label{sec:6d}
    \begin{figure}[htbp]
        \centering \includegraphics[width=0.6\textwidth]{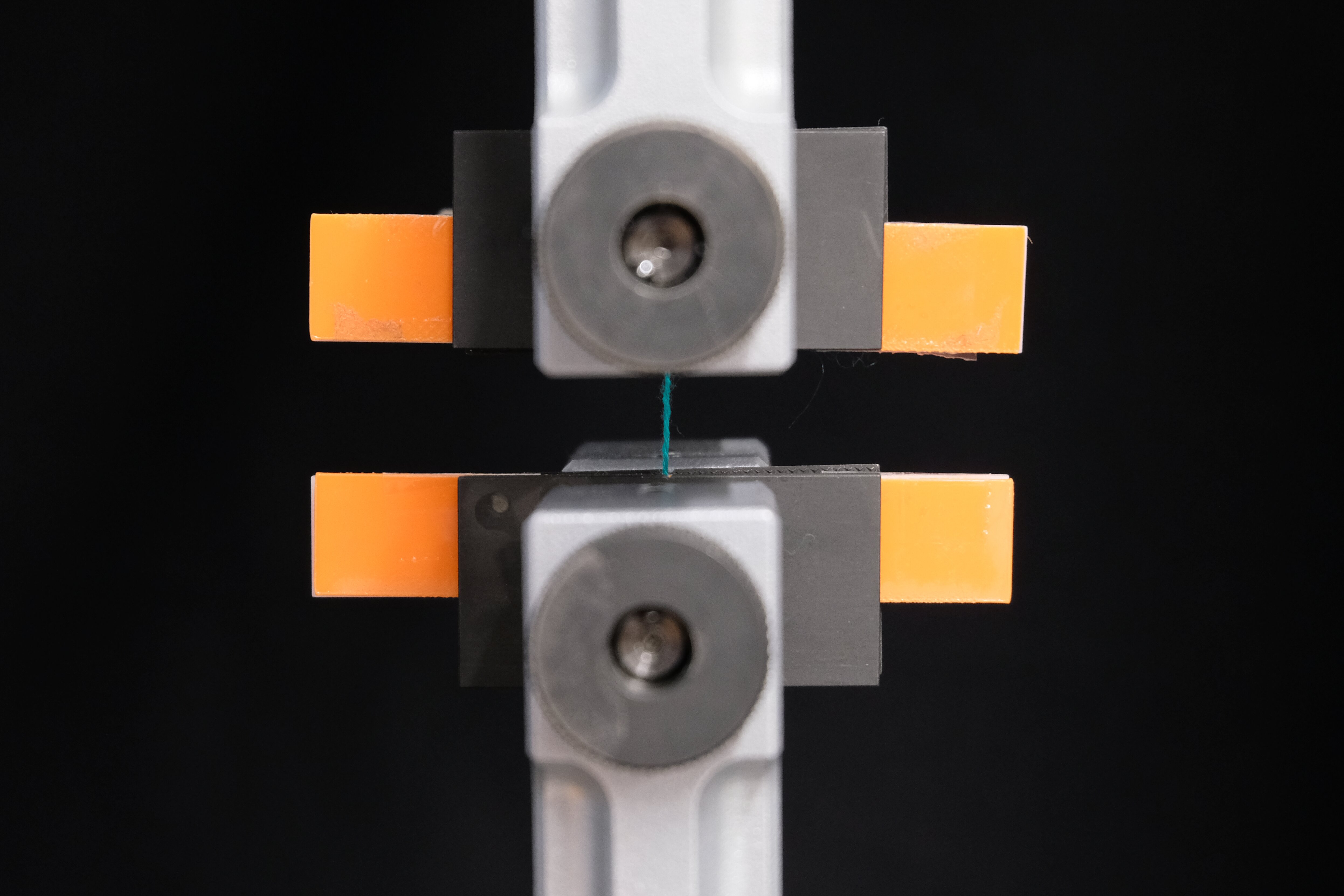}
        \caption{Tensile test on a single yarn to characterize the stretching stiffness of a single yarn.}
        \label{fig5a:yarn_pullout}
    \end{figure}
Figure \ref{fig5a:yarn_pullout} illustrates the experimental set up for a tensile test on a single yarn, in order to measure its stretching stiffness. A tensile test of a single yarn adhered to acrylic boards with instant adhesive to prevent slippage was performed on at least three different samples, with an initial fixed gauge length recorded to calculate tensile strain. All samples were tested at a rate of \SI{5}{mm/min} on a Universal Testing Machine (Instron 5566R), and tensile force $F$ and tensile strain $\epsilon$ were directly measured during loading processes. The stretching stiffness of a single yarn, defined as 
\begin{equation}
  E^s = \frac{F}{\pi r^2 \epsilon},
  \label{eq:yarn_stretch}
\end{equation}
first underwent a linear stage with \SI{79.0}{MPa} within \SI{5}{\%} tensile strain, followed by softening behaviour. In all simulations, we assumed linear elasticity and calibrated the equivalent stretching stiffness to be \SI{7.9e8}{g/(cm.s^2)}. This assumption was further confirmed by the histograms of the stretch measurements of individual yarn segments. On average, less than \SI{20}{\%} of yarn segments are stretched by more than \SI{5}{\%} even when the fabric samples are stretched by more then \SI{60}{\%}. 
    \begin{figure}[htbp]
        \centering \includegraphics[width=0.6\textwidth]{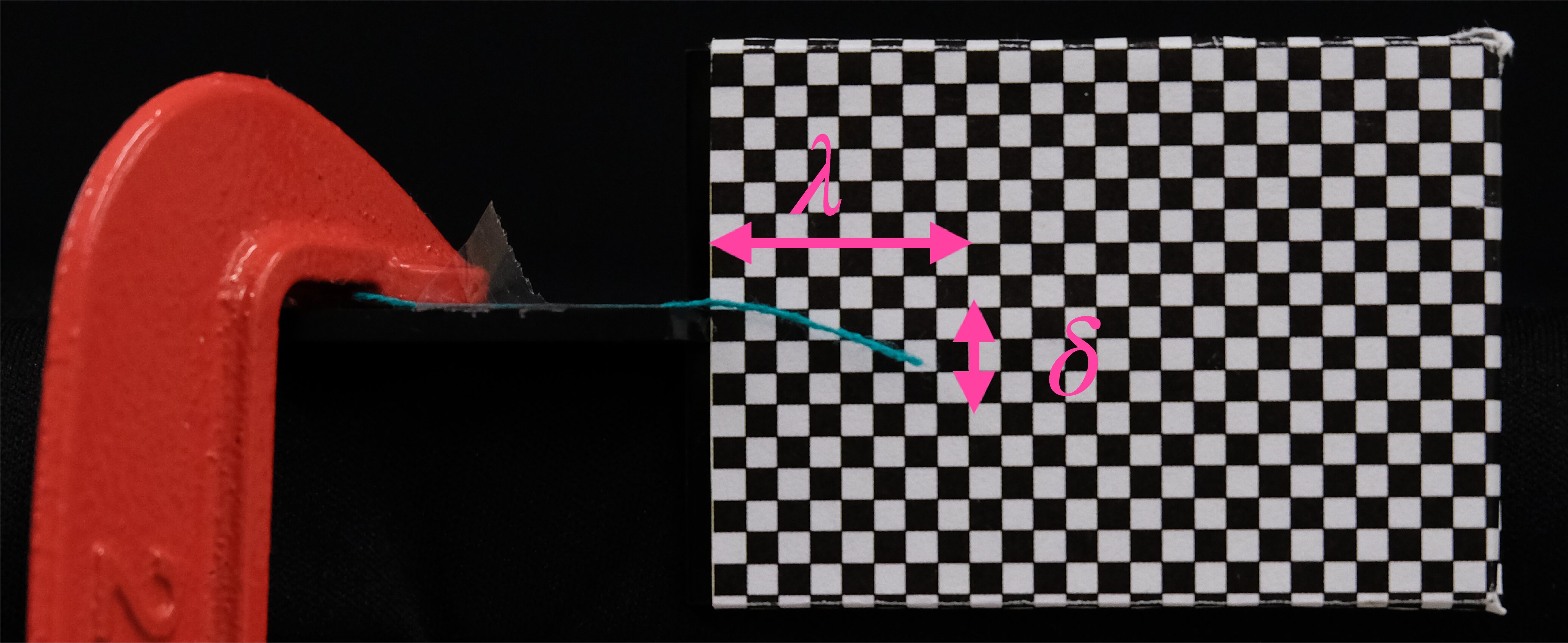}
        \caption{Bending test on a cantilevered single yarn to characterize the bending stiffness of a single yarn.}
        \label{fig5b:yarn_bend}
    \end{figure}    
Figure \ref{fig5b:yarn_bend} illustrates the experimental set up for a bending test on a single yarn cantilevered at one end with the free end consisting a horizontal length $\lambda$, and drops by a vertical displacement of $\delta$ due to gravity. We assume that gravity of the yarn acts as a distributed load $q=\pi r^2 \rho g$ and referring to Chakrabarti et al.~\cite{Chakrabarti2021}, the bending stiffness of such soft material can be determined from sets of fixed $\lambda$ and measured $\delta$ from
\begin{equation}
  E^b = \frac{q \lambda^4}{8 \delta I}.
  \label{eq:yarn_bend}
\end{equation}
For each bending test, we took high resolution photos of the cantilevered yarn segment, with pixel images on the background. Each pixel grid represents \SI{0.25}{mm} and post processing in Adobe Photoshop was done to ensure an orthogonal perspective. From a series of bending tests on three different yarn segments, we calibrated the bending stiffness of the used acrylic spun yarn to be \SI{0.249}{MPa} and its equivalent parameter in simulation is \SI{2.49e4}{g/(cm.s^2)}. Note that the measured bending stiffness differs from the stretching stiffness by several magnitudes due to hierarchical structure of the yarn.

{Table \ref{tab1:calibration} also involves several other parameter choices. The parameter $k_g$ sets the strength of a global drag that is applied to the sample via Eq.~\eqref{eq:damp_en_g}. SI Figure 3(a) demonstrates that the value of \SI{1e6}{g/s} prevents unfavourable oscillatory behavior, while not having a large effect on the stress--strain curve. Three other parameters $k$, $k_{dt}$, and $k_{dn}$ control details of the contact forces and frictional forces. SI Figures 3(b--d) demonstrate that the results are insensitive to these parameters across several orders of magnitude.}

%\item Artificial density scaling \\
%\item Global damping in order to remain in the quasi-static regime \\
%\item Other parameters: \\
%  \begin{itemize}
%      \item Contact parameters \\
%      \item $k_{dt}$ and $k_{dn}$ \\
%      \item Tolerance on the contact sphere envelope (0.03) \\
%    \end{itemize}

\subsection{Fabrication of knitted fabrics}
\label{sec:6e}
For reference we summarise the experimental protocol to fabricate and to test the fabric swatches from Sanchez et al.~\cite{Sanchez2023} here. In total, six samples for each knitted structure (jersey, garter 1 by 1, rib 1 by 1 and seed 1 by 1) were fabricated with the same settings on stitch spacing and machine tension on a Kniterate V-bed knitting machine (Kniterate, EU), and all consisting of 41 wales by 40 courses (i.e., number of stitches along the warp and the weft directions respectively). The ``knitout'' program was used to convert the Python frontend (specified by McCann et al.~\cite{McCann2016}) to Kniterate-specific machine language ``kcode'', in order to operate the knitting machine. The stitch unit arclength of each fabric $l$ was determined by 
\begin{equation}
  l = \frac{M}{\rho N_w N_c},
  \label{eq:yarn_arclength}
\end{equation}
from directly measured mass of fabricated sample $M$, fixed yarn linear density and number of wales $N_w$ and number of courses $N_c$. After fabrication, all samples were left at ambient conditions for 24-48 hours to enable material relaxation, as responding to residual stresses from the manufacturing process. Then, the dimensions of all samples were measured and approximated as those of samples in their reference states.

\subsection{Uniaxial tensile tests to characterise the knitted fabrics}
\label{sec:6f}
All knitted fabric samples were cyclically tested using Universal Testing Machines. The initial gauge lengths were recorded, in order to offset mismatch between previously recorded sample dimensions and ground-truth sample dimensions in stress-free states on the testing machines. These samples were pre-cycled to force magnitude of \SI{20}{N} at a fixed rate of \SI{10}{mm/min} during loading stage and \SI{20}{mm/min} during unloading, followed by two cycles at a rate of \SI{5}{mm/min} until reaching \SI{15}{N}. This upper bound on force was selected empirically, in order to prevent plastic deformation of the yarns within the fabric and failure (e.g., unraveling, fracture, detachment from test fixtures). For our study, we selected the mechanical responses measured after pre-cycling, as they were robustly repeated through cycles for the same fabric and the same tensile loading direction. Note that the strain ranges from experiments were smaller than those from numerics presented in this study, since the former were collected from a particular cycle of the cyclic tests, intentionally set to avoid sample failure.

\label{sec:7}

%%%% Acknowledgements %%%%%%%%%%%%%%%%%
\enlargethispage{20pt}

%\ethics{Insert ethics text here.}

%\dataccess{Insert data access text here.}
\newpage
\section*{Acknowledgment} 
This research was primarily supported by NSF through the Harvard University Materials Research Science and Engineering Center DMR-2011754. V.S. acknowledges salary support from the National Defense Science and Engineering Graduate Fellowship and the NSF under Grant No. DMR-2138020 for the MPS-Ascend program.

\section*{Author Contribution} 
C.H.R.\ and K.B.\ designed the research and supervised the project. X.D.\ and C.H.R.\ carried out analytical derivations and coded the numerical model. X.D.\ and V.S.\ designed and conducted experiments. X.D.\ performed data analysis on raw experimental data and validation of numerical model from experimental data. X.D.\ and C.H.R.\ wrote the manuscript with input from all authors.

%\funding{Insert funding text here.}

%\ack{Insert acknowledgment text here.}

%\disclaimer{Insert disclaimer text here.}

%%%% Insert bibliography here %%%%%%%%%%%
\clearpage
\bibliographystyle{unsrt}
\bibliography{citations.bib}

\begin{thebibliography}{10}

\bibitem{Poincloux2018}
Samuel Poincloux, Mokhtar Adda-Bedia, and Frédéric Lechenault.
\newblock Crackling dynamics in the mechanical response of knitted fabrics.
\newblock {\em Physical Review Letters}, 121, 7 2018.

\bibitem{Poincloux2018a}
Samuel Poincloux, Mokhtar Adda-Bedia, and Frédéric Lechenault.
\newblock Geometry and elasticity of a knitted fabric.
\newblock {\em Physical Review X}, 8, 6 2018.

\bibitem{Rout2022}
Sangram~K. Rout, Marisa~Ravena Bisram, and Jian Cao.
\newblock Methods for numerical simulation of knit based morphable structures:
  knitmorphs.
\newblock {\em Scientific Reports}, 12, 12 2022.

\bibitem{Compton2014}
Brett~G. Compton and Jennifer~A. Lewis.
\newblock 3d-printing of lightweight cellular composites.
\newblock {\em Advanced Materials}, 26(34):5930--5935, 2014.

\bibitem{zheng14}
Xiaoyu Zheng, Howon Lee, Todd~H. Weisgraber, Maxim Shusteff, Joshua DeOtte,
  Eric~B. Duoss, Joshua~D. Kuntz, Monika~M. Biener, Qi~Ge, Julie~A. Jackson,
  Sergei~O. Kucheyev, Nicholas~X. Fang, and Christopher~M. Spadaccini.
\newblock Ultralight, ultrastiff mechanical metamaterials.
\newblock {\em Science}, 344(6190):1373--1377, 2014.

\bibitem{Jiang2016}
Yanhui Jiang and Qiming Wang.
\newblock Highly-stretchable {3D}-architected mechanical metamaterials.
\newblock {\em Scientific Reports}, 6, 9 2016.

\bibitem{Moestopo2023}
Widianto~P Moestopo, Sammy Shaker, Weiting Deng, and Julia~R Greer.
\newblock Knots are not for naught: Design, properties, and topology of
  hierarchical intertwined microarchitected materials.
\newblock {\em Science Advances}, 2023.

\bibitem{Gladman2016}
A.~Sydney Gladman, Elisabetta~A. Matsumoto, Ralph~G. Nuzzo, L.~Mahadevan, and
  Jennifer~A. Lewis.
\newblock Biomimetic {4D} printing.
\newblock {\em Nature Materials}, 15:413--418, 4 2016.

\bibitem{Ma2017}
Yinji Ma, Xue Feng, John~A Rogers, Yonggang Huang, and Yihui Zhang.
\newblock Design and application of {'J-shaped'} stress-strain behavior in
  stretchable electronics: A review.
\newblock {\em Lab on a Chip}, 17:1689--1704, 2017.

\bibitem{Nepal2023}
Dhriti Nepal, Saewon Kang, Katarina~M. Adstedt, Krishan Kanhaiya, Michael~R.
  Bockstaller, L.~Catherine Brinson, Markus~J. Buehler, Peter~V. Coveney,
  Kaushik Dayal, Jaafar~A. El-Awady, Luke~C. Henderson, David~L. Kaplan, Sinan
  Keten, Nicholas~A. Kotov, George~C. Schatz, Silvia Vignolini, Fritz Vollrath,
  Yusu Wang, Boris~I. Yakobson, Vladimir~V. Tsukruk, and Hendrik Heinz.
\newblock Hierarchically structured bioinspired nanocomposites.
\newblock {\em Nature Materials}, 22:18--35, 1 2023.

\bibitem{Mistry2023}
Yash Mistry, Oliver Weeger, Swapnil Morankar, Mandar Shinde, Siying Liu,
  Nikhilesh Chawla, Xiangfan Chen, Clint~A. Penick, and Dhruv Bhate.
\newblock Bio-inspired selective nodal decoupling for ultra-compliant
  interwoven lattices.
\newblock {\em Communications Materials}, 4, 12 2023.

\bibitem{Polygerinos2015}
Panagiotis Polygerinos, Zheng Wang, Kevin~C Galloway, Robert~J Wood, and
  Conor~J Walsh.
\newblock Soft robotic glove for combined assistance and at-home
  rehabilitation.
\newblock {\em Robotics and Autonomous Systems}, 2015.

\bibitem{Cappello2018}
Leonardo Cappello, Kevin~C Galloway, Siddharth Sanan, Diana~A Wagner, Rachael
  Granberry, Sven Engelhardt, Florian~L Haufe, Jeffrey~D Peisner, and Conor~J
  Walsh.
\newblock Exploiting textile mechanical anisotropy for fabric-based pneumatic
  actuators.
\newblock {\em Soft robotics}, 5(5):662--674, 2018.

\bibitem{lee2018knit}
Seulah Lee, Myoung-Ok Kim, Taeho Kang, Junho Park, and Youngjin Choi.
\newblock Knit band sensor for myoelectric control of surface emg-based
  prosthetic hand.
\newblock {\em IEEE Sensors Journal}, 18(20):8578--8586, 2018.

\bibitem{Granberry2019}
Rachael Granberry, Kevin Eschen, Brad Holschuh, and Julianna Abel.
\newblock Functionally graded knitted actuators with {NiTi-based} shape memory
  alloys for topographically self-fitting wearables.
\newblock {\em Advanced Materials Technologies}, 4, 11 2019.

\bibitem{fan2020machine}
Wenjing Fan, Qiang He, Keyu Meng, Xulong Tan, Zhihao Zhou, Gaoqiang Zhang, Jin
  Yang, and Zhong~Lin Wang.
\newblock Machine-knitted washable sensor array textile for precise epidermal
  physiological signal monitoring.
\newblock {\em Science Advances}, 6(11), 2020.

\bibitem{Wicaksono2020}
Irmandy Wicaksono, Carson Tucker, Tao Sun, Cesar Guerrero, Clare Liu, Wesley
  Woo, Eric Pence, and Canan Dagdeviren.
\newblock A tailored, electronic textile conformable suit for large-scale
  spatiotemporal physiological sensing in vivo.
\newblock {\em npj Flexible Electronics}, 4:5, 4 2020.

\bibitem{Connolly2019}
Fionnuala Connolly, Diana~A. Wagner, Conor~J. Walsh, and Katia Bertoldi.
\newblock Sew-free anisotropic textile composites for rapid design and
  manufacturing of soft wearable robots.
\newblock {\em Extreme Mechanics Letters}, 27:52--58, 2 2019.

\bibitem{Sanchez2021}
Vanessa Sanchez, Conor~J. Walsh, and Robert~J. Wood.
\newblock Textile technology for soft robotic and autonomous garments.
\newblock {\em Advanced Functional Materials}, 31, 2 2021.

\bibitem{Ahlquist2017}
Sean Ahlquist, Wes McGee, and Shahida Sharmin.
\newblock Pneumaknit: Actuated architectures through wale-and course-wise
  tubular knit-constrained pneumatic systems.
\newblock In {\em Disciplines \& disruption: Proceedings of the 37th annual
  conference of the association for computer aided design in architecture},
  pages 38--51. ACADIA Cambridge, MA, 2017.

\bibitem{Luo2022}
Yiyue Luo, Kui Wu, Andrew Spielberg, Michael Foshey, Daniela Rus, Tomás
  Palacios, and Wojciech Matusik.
\newblock Digital fabrication of pneumatic actuators with integrated sensing by
  machine knitting.
\newblock {\em ACM Proceedings of Conference on Human Factors in Computing
  Systems}, 4 2022.

\bibitem{Sanchez2023}
Vanessa Sanchez, Kausalya Mahadevan, Gabrielle Ohlson, Moritz~A. Graule,
  Michelle~C. Yuen, Clark~B. Teeple, James~C. Weaver, James McCann, Katia
  Bertoldi, and Robert~J. Wood.
\newblock {3D} knitting for pneumatic soft robotics.
\newblock {\em Advanced Functional Materials}, 2023.

\bibitem{Abel2013}
Julianna Abel, Jonathan Luntz, and Diann Brei.
\newblock Hierarchical architecture of active knits.
\newblock {\em Smart Materials and Structures}, 22, 12 2013.

\bibitem{Han2017}
Min-Woo Han and Sung-Hoon Ahn.
\newblock Blooming knit flowers: Loop-linked soft morphing structures for soft
  robotics.
\newblock {\em Advanced Materials}, 29(13), 2017.

\bibitem{Albaugh2019}
Lea Albaugh, Lining Yao, and Scott Hudson.
\newblock Digital fabrication of soft actuated objects by machine knitting.
\newblock {\em Extended Abstracts of the 2019 CHI Conference on Human Factors
  in Computing Systems}, 2019.

\bibitem{Luo2021}
Yiyue Luo, Yunzhu Li, Pratyusha Sharma, Wan Shou, Kui Wu, Michael Foshey,
  Beichen Li, Tomás Palacios, Antonio Torralba, and Wojciech Matusik.
\newblock Learning human–environment interactions using conformal tactile
  textiles.
\newblock {\em Nature Electronics}, 4:193--201, 2021.

\bibitem{Wicaksono2017}
Irmandy Wicaksono and Joseph~A Paradiso.
\newblock Fabrickeyboard: multimodal textile sensate media as an expressive and
  deformable musical interface.
\newblock {\em New Interfaces for Musical Expression}, 17:348--353, 2017.

\bibitem{Pei2019}
Zeguang Pei, Xiangzhang Xiong, Jian He, and Yan Zhang.
\newblock Highly stretchable and durable conductive knitted fabrics for the
  skins of soft robots.
\newblock {\em Soft Robotics}, 6:687--700, 12 2019.

\bibitem{Peirce1947}
F.T. Peirce.
\newblock Geometrical principles applicable to the design of functional
  fabrics.
\newblock {\em Textile Research Journal}, 17(3):123--147, 1947.

\bibitem{Leaf1955}
G.~A.~V. Leaf and A.~Glaskin.
\newblock 43—the geometry of a plain knitted loop.
\newblock {\em Journal of the Textile Institute Transactions}, 46(9):587--605,
  1955.

\bibitem{Munden1959}
D.~L. Munden.
\newblock 26—the geometry and dimensional properties of plain-knit fabrics.
\newblock {\em Journal of the Textile Institute Transactions}, 50(7):448--471,
  1959.

\bibitem{Bergou2008}
Miklós Bergou, Max Wardetzky, Stephen Robinson, and Basile Audoly.
\newblock Discrete elastic rods.
\newblock {\em ACM Transactions on Graphics}, 21:1--12, 2008.

\bibitem{Kaldor2008}
Jonathan~M. Kaldor, Doug~L. James, and Steve Marschner.
\newblock Simulating knitted cloth at the yarn level.
\newblock {\em ACM Transactions on Graphics}, 27:1--9, 8 2008.

\bibitem{Cirio2014}
Gabriel Cirio, Jorge Lopez-Moreno, David Miraut, and Miguel~A. Otaduy.
\newblock Yarn-level simulation of woven cloth.
\newblock {\em ACM Transactions on Graphics}, 33, 11 2014.

\bibitem{Liu2017}
Dani Liu, Daniel Christe, Bahareh Shakibajahromi, Chelsea Knittel, Nestor
  Castaneda, David Breen, Genevieve Dion, and Antonios Kontsos.
\newblock On the role of material architecture in the mechanical behavior of
  knitted textiles.
\newblock {\em International Journal of Solids and Structures}, 109:101--111,
  2017.

\bibitem{Leaf2018}
Jonathan Leaf, Rundong Wu, Eston Schweickart, Doug~L. James, and Steve
  Marschner.
\newblock Interactive design of periodic yarn-level cloth patterns.
\newblock {\em ACM Transactions on Graphics}, 37:1--15, 12 2018.

\bibitem{Wu2020}
Liwei Wu, Feng Zhao, Junbo Xie, Xianyan Wu, Qian Jiang, and Jia-Horng Lin.
\newblock The deformation behaviors and mechanism of weft knitted fabric based
  on micro-scale virtual fiber model.
\newblock {\em International Journal of Mechanical Sciences}, 187, 2020.

\bibitem{Terzopoulos1987}
Demetri Terzopoulos, John Platt, Alan Barr, and Kurt Fleischert.
\newblock Elastically deformable models.
\newblock {\em ACM SIGGRAPH Computer Graphics}, 21:205--214, 1987.

\bibitem{Baraff1998}
David Baraff and Andrew Witkin.
\newblock Large steps in cloth simulation.
\newblock {\em ACM SIGGRAPH 1998 Proceedings of the 25th Annual Conference on
  Computer Graphics and Interactive Techniques}, 1998.

\bibitem{Breen1994}
David~E. Breen, Donald~H. House, and Michael~J. Wozny.
\newblock Predicting the drape of woven cloth using interacting particles.
\newblock {\em ACM SIGGRAPH 1994 Proceedings of the 21st Annual Conference on
  Computer Graphics and Interactive Techniques}, 1994.

\bibitem{Yeoman2010}
Mark~S. Yeoman, Daya Reddy, Hellmut~C. Bowles, Deon Bezuidenhout, Peter Zilla,
  and Thomas Franz.
\newblock A constitutive model for the warp-weft coupled non-linear behavior of
  knitted biomedical textiles.
\newblock {\em Biomaterials}, 31:8484--8493, 11 2010.

\bibitem{NarainArminSamiiJamesO2012}
Rahul F Narain Armin Samii~James O, Images copyright Rahul~Narain, Armin Samii,
  and James~F O.
\newblock Adaptive anisotropic remeshing for cloth simulation.
\newblock {\em ACM Transactions on Graphics}, 31:1--10, 2012.

\bibitem{Yuksel2012}
Cem Yuksel, Jonathan~M Kaldor, Doug~L James, and Steve Marschner.
\newblock Stitch meshes for modeling knitted clothing with yarn-level detail.
\newblock {\em ACM Transactions on Graphics}, 31:1--12, 2012.

\bibitem{Dinh2018}
Tien~D. Dinh, Oliver Weeger, Sawako Kaijima, and Sai-Kit Yeung.
\newblock Prediction of mechanical properties of knitted fabrics under tensile
  and shear loading: Mesoscale analysis using representative unit cells and its
  validation.
\newblock {\em Composites Part B: Engineering}, 148:81--92, 2018.

\bibitem{Weeger2018}
Oliver Weeger, Amir~Hosein Sakhaei, Ying~Yi Tan, Yu~Han Quek, Tat~Lin Lee,
  Sai~Kit Yeung, Sawako Kaijima, and Martin~L. Dunn.
\newblock Nonlinear multi-scale modelling, simulation and validation of 3d
  knitted textiles.
\newblock {\em Applied Composite Materials}, 25:797--810, 8 2018.

\bibitem{Sperl2021}
Georg Sperl, Rahul Narain, and Chris Wojtan.
\newblock Mechanics-aware deformation of yarn pattern geometry.
\newblock {\em ACM Transactions on Graphics}, 40:1--11, 7 2021.

\bibitem{Martinez2021}
Jorge~Llinares Berenguer, Pablo Diaz-García, and Pau~Miró Martinez.
\newblock Determining the loop length during knitting and dyeing processes.
\newblock {\em Textile Research Journal}, 91(1-2):188--199, 2021.

\bibitem{Allan1983}
S.~Allan Heap, Peter~F. Greenwood, Robert~D. Leah, James~T. Eaton, Jill~C.
  Stevens, and Pauline Keher.
\newblock Prediction of finished weight and shrinkage of cotton knits— the
  starfish project: Part i: Introduction and general overview.
\newblock {\em Textile Research Journal}, 53(2):109--119, 1983.

\bibitem{Amreeva2007}
Gulmira Amreeva and Arif Kurbak.
\newblock Experimental studies on the dimensional properties of half milano and
  milano rib fabrics.
\newblock {\em Textile Research Journal}, 77(3):151--160, 2007.

\bibitem{Wei2011}
Linna Wei and Lihua Chen.
\newblock Research on influence of pre-tension on elongation percentage at
  specified load for knits.
\newblock {\em Advanced Textile Materials, Part 1}, 2011.

\bibitem{Cao2008}
J.~Cao, R.~Akkerman, P.~Boisse, J.~Chen, H.~S. Cheng, E.~F. de~Graaf, J.~L.
  Gorczyca, P.~Harrison, G.~Hivet, J.~Launay, W.~Lee, L.~Liu, S.~V. Lomov,
  A.~Long, E.~de~Luycker, F.~Morestin, J.~Padvoiskis, X.~Q. Peng, J.~Sherwood,
  Tz~Stoilova, X.~M. Tao, I.~Verpoest, A.~Willems, J.~Wiggers, T.~X. Yu, and
  B.~Zhu.
\newblock Characterization of mechanical behavior of woven fabrics:
  Experimental methods and benchmark results.
\newblock {\em Composites Part A: Applied Science and Manufacturing},
  39:1037--1053, 6 2008.

\bibitem{Markande2020}
Shashank~G Markande and Elisabetta~A Matsumoto.
\newblock Knotty knits are tangles on tori.
\newblock {\em arXiv: 2002.01497}, 2020.

\bibitem{barchiesi19}
Emilio Barchiesi, Mario Spagnuolo, and Luca Placidi.
\newblock Mechanical metamaterials: a state of the art.
\newblock {\em Mathematics and Mechanics of Solids}, 24(1):212--234, 2019.

\bibitem{florijn14}
Bastiaan Florijn, Corentin Coulais, and Martin van Hecke.
\newblock Programmable mechanical metamaterials.
\newblock {\em Phys. Rev. Lett.}, 113:175503, 10 2014.

\bibitem{grima13}
Joseph~N. Grima, Roberto Caruana-Gauci, Miros{\l}aw~R Dudek, Krzysztof~W.
  Wojciechowski, and Ruben Gatt.
\newblock Smart metamaterials with tunable auxetic and other properties.
\newblock {\em Smart Materials and Structures}, 22(8), 2013.

\bibitem{duoss19}
Eric~B. Duoss, Todd~H. Weisgraber, Keith Hearon, Cheng Zhu, Ward Small~IV,
  Thomas~R. Metz, John~J. Vericella, Holly~D. Barth, Joshua~D. Kuntz, Robert~S.
  Maxwell, Christopher~M. Spadaccini, and Thomas~S. Wilson.
\newblock Three-dimensional printing of elastomeric, cellular architectures
  with negative stiffness.
\newblock {\em Advanced Functional Materials}, 24(31):4905--4913, 2014.

\bibitem{yuan19}
Shangqin Yuan, Chee~Kai Chua, and Kun Zhou.
\newblock 3d-printed mechanical metamaterials with high energy absorption.
\newblock {\em Advanced Materials Technologies}, 4(3), 2019.

\bibitem{medina20}
Eder Medina, Patrick~E. Farrell, Katia Bertoldi, and Chris~H. Rycroft.
\newblock Navigating the landscape of nonlinear mechanical metamaterials for
  advanced programmability.
\newblock {\em Phys. Rev. B}, 101, 2 2020.

\bibitem{chan20}
Yu-Chin Chan, Faez Ahmed, Liwei Wang, and Wei Chen.
\newblock {METASET}: {Exploring} shape and property spaces for data-driven
  metamaterials design.
\newblock {\em Journal of Mechanical Design}, 143(3), 11 2020.

\bibitem{mao20}
Yunwei Mao, Qi~He, and Xuanhe Zhao.
\newblock Designing complex architectured materials with generative adversarial
  networks.
\newblock {\em Science Advances}, 6(17), 2020.

\bibitem{deng22}
Bolei Deng, Ahmad Zareei, Xiaoxiao Ding, James~C. Weaver, Chris~H. Rycroft, and
  Katia Bertoldi.
\newblock Inverse design of mechanical metamaterials with target nonlinear
  response via a neural accelerated evolution strategy.
\newblock {\em Advanced Materials}, 34(41), 2022.

\bibitem{medina23}
Eder Medina, Chris~H. Rycroft, and Katia Bertoldi.
\newblock Nonlinear shape optimization of flexible mechanical metamaterials.
\newblock {\em Extreme Mechanics Letters}, 61, 2023.

\bibitem{suli_textbook}
Endre S\"uli and David~F. Mayers.
\newblock {\em An Introduction to Numerical Analysis}.
\newblock Cambridge University Press, 2003.

\bibitem{Vassiliadis2007}
Savvas Vassiliadis, Argyro Kallivretaki, and Christopher Provatidis.
\newblock Mechanical simulation of the plain weft knitted fabrics.
\newblock {\em International Journal of Clothing Science and Technology},
  19:109--130, 3 2007.

\bibitem{Syerk2012}
Elena Syerko, Sébastien Comas-Cardona, and Christophe Binetruy.
\newblock Models of mechanical properties/behavior of dry fibrous materials at
  various scales in bending and tension: A review.
\newblock {\em Composites Part A: Applied Science and Manufacturing},
  43(8):1365--1388, 2012.

\bibitem{Spencer2001}
David~J. Spencer.
\newblock {\em Knitting Technology - A Comprehensive Handbook and Practical
  Guide (3rd Edition)}.
\newblock Woodhead Publishing Series in Textiles. Woodhead Publishing,
  Cambridge, 2001.

\bibitem{Piuze2011}
Emmanuel Piuze, Paul~G. Kry, and Kaleem Siddiqi.
\newblock Generalized helicoids for modeling hair geometry.
\newblock {\em Computer Graphics Forum}, 30:247--256, 2011.

\bibitem{Wadekar2020}
Paras Wadekar, Prateek Goel, Chelsea Amanatides, Genevieve Dion, Randall~D.
  Kamien, and David~E. Breen.
\newblock Geometric modeling of knitted fabrics using helicoid scaffolds.
\newblock {\em Journal of Engineered Fibers and Fabrics}, 15, 2020.

\bibitem{Eltahan2016}
Eman Abd~Elzaher Eltahan, Mohamed Sultan, and Abou-Bakr Mito.
\newblock Determination of loop length, tightness factor and porosity of single
  jersey knitted fabric.
\newblock {\em Alexandria Engineering Journal}, 55:851--856, 6 2016.

\bibitem{Meyers2013}
Marc~André Meyers, Joanna McKittrick, and Po-Yu Chen.
\newblock Structural biological materials: Critical mechanics-materials
  connections.
\newblock {\em Science}, 339(6121):773--779, 2013.

\bibitem{Barthelat2016}
Francois Barthelat, Zhen Yin, and Markus~J. Buehler.
\newblock Structure and mechanics of interfaces in biological materials.
\newblock {\em Nature Reviews Materials}, 1, 3 2016.

\bibitem{Oosten2019}
Anne~S.G. van Oosten, Xingyu Chen, Li~Kang Chin, Katrina Cruz, Alison~E.
  Patteson, Katarzyna Pogoda, Vivek~B. Shenoy, and Paul~A. Janmey.
\newblock Emergence of tissue-like mechanics from fibrous networks confined by
  close-packed cells.
\newblock {\em Nature}, 573:96--101, 9 2019.

\bibitem{Zhang2021}
Yao Zhang, Jingyi Yu, Xuan Wang, Daniel~M. Durachko, Sulin Zhang, and Daniel~J.
  Cosgrove.
\newblock Molecular insights into the complex mechanics of plant epidermal cell
  walls.
\newblock {\em Science}, 372(6543):706--711, 2021.

\bibitem{dagum98}
L.~Dagum and R.~Menon.
\newblock {OpenMP: an industry standard API for shared-memory programming}.
\newblock {\em IEEE Computational Science and Engineering}, 5(1):46--55, 1998.

\bibitem{hairer93}
E.~Hairer, S.~P. N{\o}rsett, and G.~Wanner.
\newblock {\em Solving Ordinary Differential Equations {I}: {N}onstiff
  Problems}.
\newblock Springer, Berlin, 1993.

\bibitem{andrejevic22}
Jovana Andrejevic and Chris~H. Rycroft.
\newblock Simulation of crumpled sheets via alternating quasistatic and dynamic
  representations.
\newblock {\em Journal of Computational Physics}, 471:111607, 2022.

\bibitem{peskin02}
Charles~S. Peskin.
\newblock The immersed boundary method.
\newblock {\em Acta Numerica}, 11:479--517, 1 2002.

\bibitem{fai18}
Thomas~G. Fai and Chris~H. Rycroft.
\newblock Lubricated immersed boundary method in two dimensions.
\newblock {\em Journal of Computational Physics}, 356:319--339, 2018.

\bibitem{Chakrabarti2021}
Aditi Chakrabarti, Salem Al-Mosleh, and L.~Mahadevan.
\newblock Instabilities and patterns in a submerged jelling jet.
\newblock {\em Soft Matter}, 17:9745--9754, 11 2021.

\bibitem{McCann2016}
James McCann, Lea Albaugh, Vidya Narayanan, April Grow, Wojciech Matusik,
  Jennifer Mankoff, and Jessica Hodgins.
\newblock A compiler for {3D} machine knitting.
\newblock {\em ACM Trans. Graph.}, 35(4), 7 2016.

\end{thebibliography}

%%%% SI %%%%%%%%%%%%%%%%%%%%%%%%
%\section{SI}
\clearpage
\label{sec:si}
% Supplementary Material
%\documentclass[%
 %reprint,
%superscriptaddress,
%groupedaddress,
%unsortedaddress,
%runinaddress,
%frontmatterverbose,
%preprint,
%preprintnumbers,
%nofootinbib,
%nobibnotes,
%bibnotes,
 %amsmath,amssymb,
 %aps,
%pra,
%prb,
%rmp,
%prstab,
%prstper,
%floatfix,
%]{revtex4-2}

%\usepackage{graphicx}% Include figure files
%\usepackage{dcolumn}% Align table columns on decimal point
%\usepackage{bm}% bold math
%\usepackage{mathtools} % dcases
%\usepackage{microtype}
%\usepackage{siunitx}
%\usepackage{color}

\def\bibsection{\section*{Supplementary References}}
%\usepackage{caption}
%\captionsetup{justification=raggedright,singlelinecheck=false,format=hang}
\renewcommand{\figurename}{Supplementary Figure}

%\definecolor{webgreen}{rgb}{0,.5,0}
%\usepackage[colorlinks=true,citecolor=webgreen]{hyperref}

\definecolor{violet}{rgb}{0.4,0.2,0.8}
\newcommand{\CHR}[1]{\textcolor{violet}{(CHR: #1)}}
\definecolor{orange}{rgb}{0.9,0.5,0.2}
\newcommand{\JA}[1]{\textcolor{orange}{(JA: #1)}}

%\usepackage{hyperref}% add hypertext capabilities
%\usepackage[mathlines]{lineno}% Enable numbering of text and display math
%\linenumbers\relax % Commence numbering lines

%\usepackage[showframe,%Uncomment any one of the following lines to test
%%scale=0.7, marginratio={1:1, 2:3}, ignoreall,% default settings
%%text={7in,10in},centering,
%%margin=1.5in,
%%total={6.5in,8.75in}, top=1.2in, left=0.9in, includefoot,
%%height=10in,a5paper,hmargin={3cm,0.8in},
%]{geometry}

%\begin{document}
%\preprint{APS/123-QED}

\setcounter{section}{0}
\setcounter{equation}{0}
\setcounter{figure}{0}
\setcounter{table}{0}
\setcounter{page}{1}
\makeatletter
%\renewcommand{\theequation}{S\arabic{equation}}
%\renewcommand{\thefigure}{S\arabic{figure}}
%\renewcommand{\bibnumfmt}[1]{[S#1]}
%\renewcommand{\citenumfont}[1]{S#1}

%\onecolumngrid
\begin{center}
    \large\textbf{Unravelling the mechanics of knitted fabrics through hierarchical geometric representation} \\
    \vspace{\baselineskip}
    \normalsize Xiaoxiao Ding,\textsuperscript{1,2} Vanessa Sanchez,\textsuperscript{1,3} Katia Bertoldi,\textsuperscript{1} and Chris H.~Rycroft\textsuperscript{2,4} \\
    \textsuperscript{1}\textit{John A.~Paulson School of Engineering and Applied Sciences,} \\
    \textit{Harvard University, Cambridge, MA 02138, USA} \\
    \textsuperscript{2}\textit{Department of Mathematics, University of Wisconsin--Madison, Madison, WI 53706, USA} \\
    \textsuperscript{3}\textit{Department of Chemical Engineering, Stanford University, 443 Via Ortega, Stanford, CA 94305, USA} \\
    \textsuperscript{4}\textit{Computational Research Division, Lawrence Berkeley Laboratory, 1 Cyclotron Road, Berkeley, CA 94720, USA}
\end{center}

\section*{Supplementary Figure 1}
\begin{figure}[htb!]
    \centering \includegraphics[width=1\textwidth]{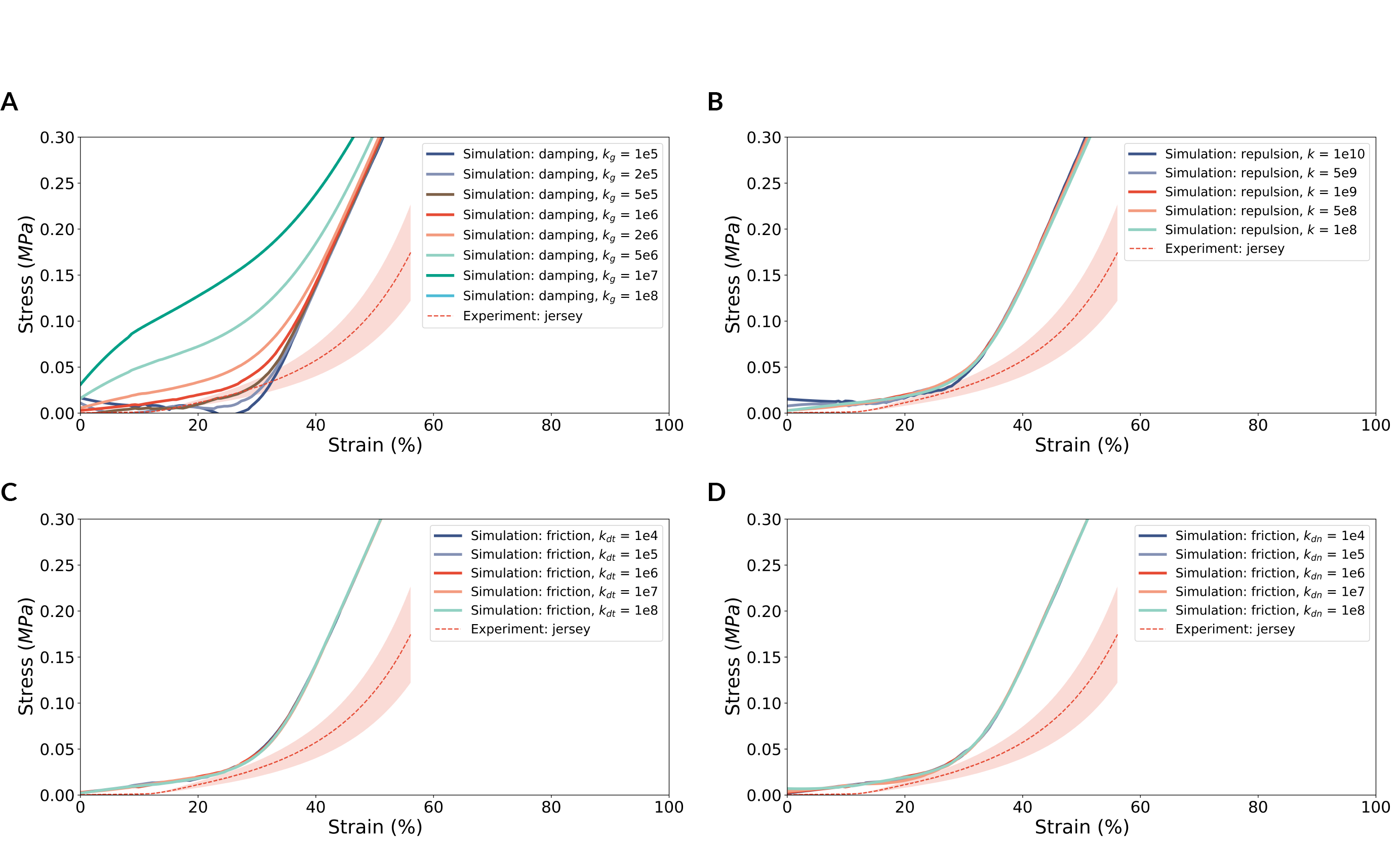}
    \caption{Parametric study on the effects on a jersey knitted sample subjected to external tension along fabric warp: (A) from damping mass, $k_g$ (units of \si{g/s}) (B) from contact repulsion constant, $k$, (units of \si{g/cm.s^2}) (C) from tangential frictional coefficient $k_{dt}$ (units of \si{g/cm^2.s}) and (D) from normal frictional coefficient $k_{dn}$ (units of \si{g/cm^2.s}).}
    \label{fig_re1:params}
\end{figure}

\clearpage
\section*{Supplementary Figure 2}
\begin{figure}[!htb]
    \centering \includegraphics[width=1\textwidth]{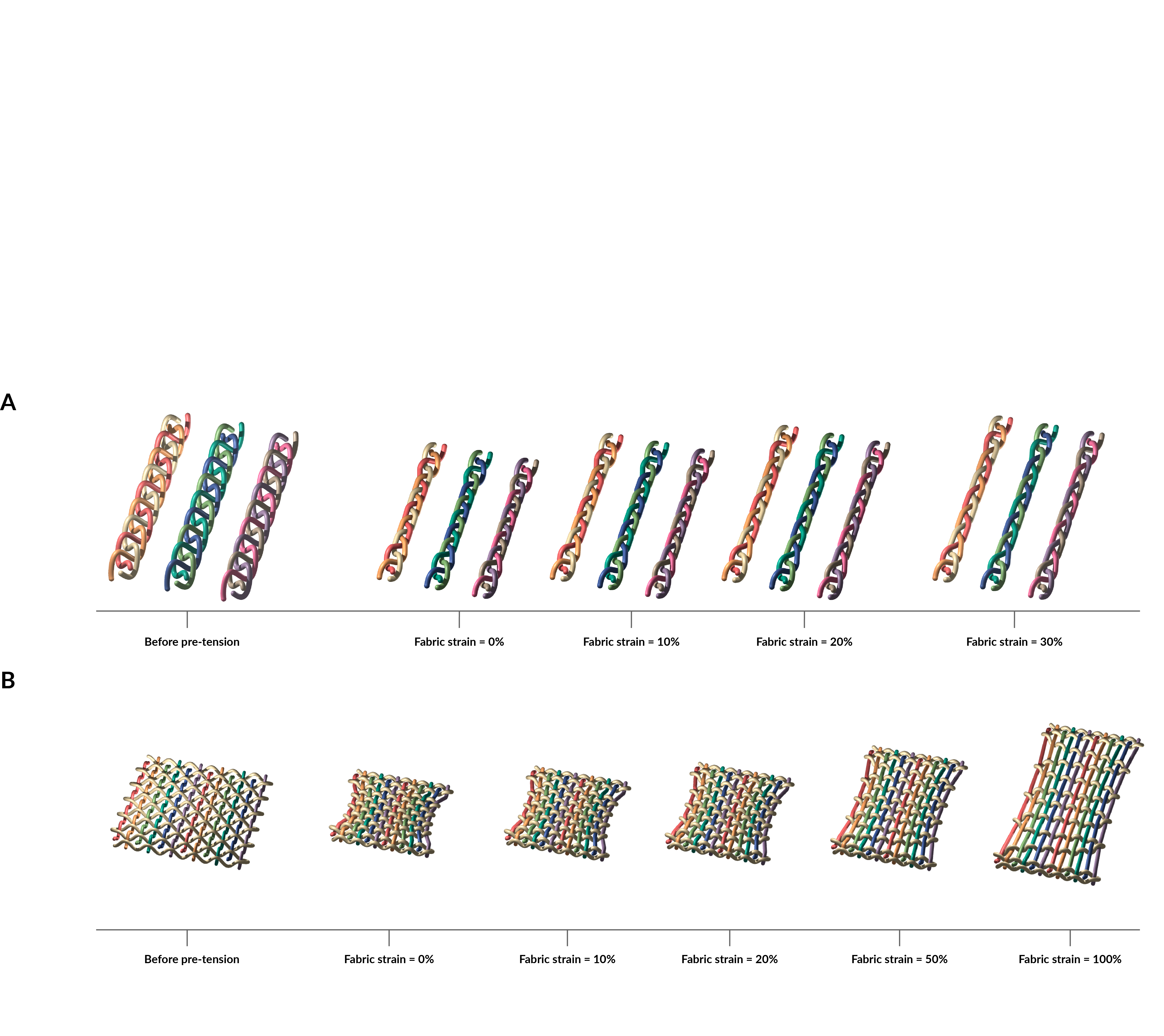}
    \caption{(A) Demonstration of the capability of our model to simulate three braided structures each consisting of three yarns with open ends. (B) Simulation of a woven structure consisting of 12 yarns with open ends in parallel and one additional yarn to connect the ends of stress-free boundaries. In both (A) and (B) each yarn is specified with the same material property. However, we colour-code each yarn in a different colour, as to show the capability of specifying different yarn-wise material properties if one may wish to design such braided and woven structures compromising stiff and soft yarns into the same textile structure.}
    \label{figsi:braid_woven}
\end{figure}

\clearpage
\section*{Supplementary Figure 3}
\begin{figure}[htb!]
        \centering
        \includegraphics[width=1\textwidth]{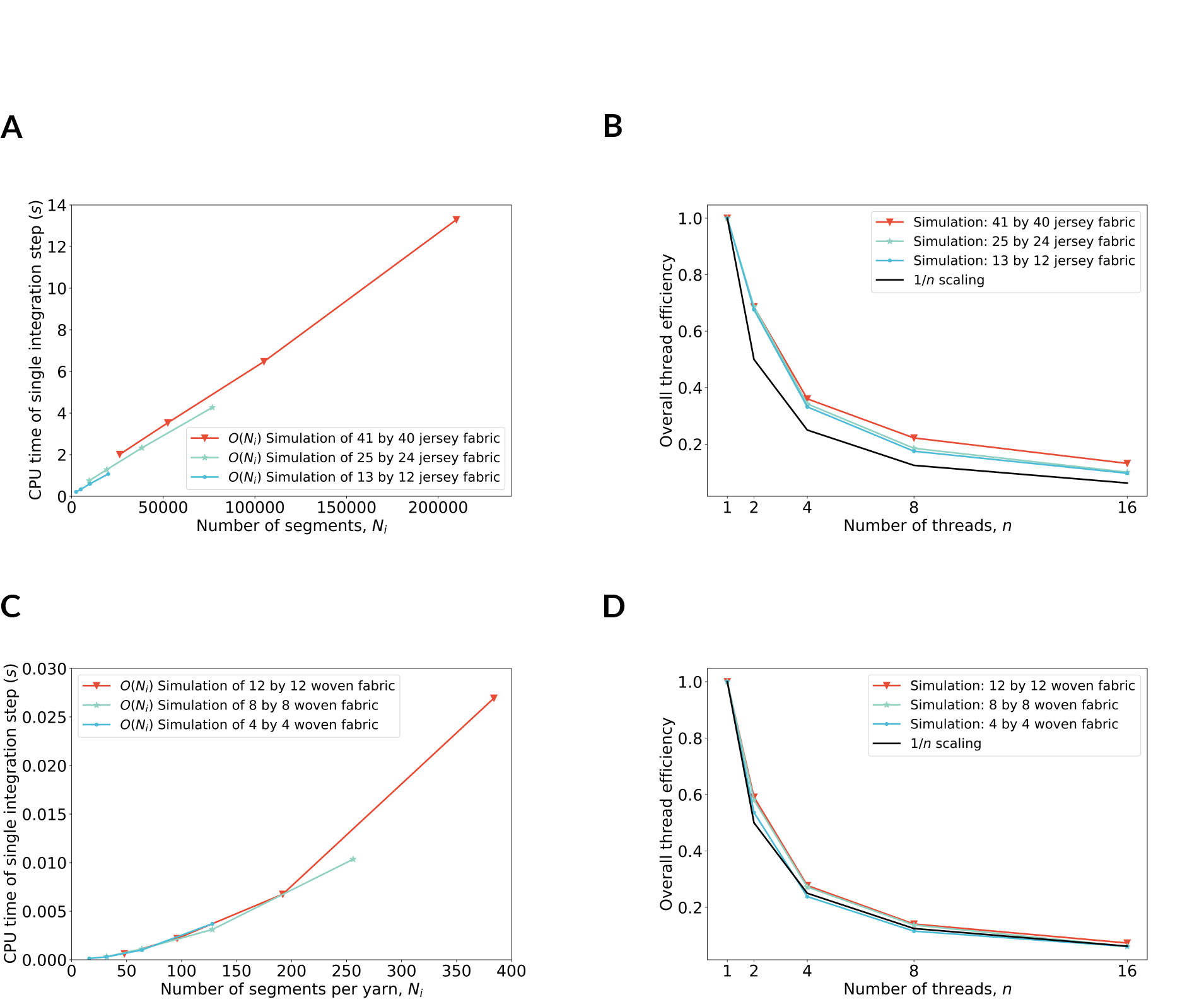}
        \caption{Study on computational efficiency on textiles consisting of single yarn (jersey knitted fabric) and multiple yarns (woven fabric): (A) CPU time of testing jersey knitted fabric subject to tension along fabric warp scales linearly across multiple system sizes, where $N_i$ is the total number of segments; (B) Overall thread efficiency of testing jersey knitted fabric subject to tension along fabric warp shows advantage of multi-threading; (C) CPU time of testing woven fabric (Supplemental Fig.~\ref{figsi:braid_woven}) subject to tension along fabric warp scales almost linearly across multiple system sizes, where $N_i$ is the number of segments per yarn; (D) Overall thread efficiency of testing woven fabric subject to tension along fabric warp shows advantage of multi-threading.}
        \label{fig_re2:time}
\end{figure}

%\end{document}

\end{document}